\documentclass[twocolumn,amsmath,amssymb,nofootinbib,tighten,floatfix,prb]{revtex4}

\usepackage{graphicx}
\usepackage{dcolumn} 
\usepackage{bm, bbm}      
\usepackage{curves}
\usepackage{epic}
\usepackage{wasysym}
\usepackage{epsf, times}
\usepackage{subfigure}
\usepackage{eufrak}
\usepackage{amsthm}

\newcommand{\beq}{\begin{equation}}
\newcommand{\eeq}{\end{equation}}
\newcommand{\bea}{\begin{eqnarray}}
\newcommand{\eea}{\end{eqnarray}}

\def\openone{\leavevmode\hbox{\small1\kern-4.2pt\normalsize1}}
\def\A{\mathcal{A}}
\def\B{\mathcal{B}}
\def\P{\mathcal{P}}

\begin{document}


\title{
Entanglement and topological entropy of the toric code at finite temperature
      }

\author{
Claudio Castelnovo$^1$ 
and 
Claudio Chamon$^2$
       }
\affiliation{
$^1$ 
Rudolf Peierls Centre for Theoretical Physics, 
University of Oxford, Oxford, OX1 3NP, UK
\\ 
$^2$ 
Physics Department, Boston University, Boston, MA 02215, USA
            }

\date{\today}

\begin{abstract}
We calculate exactly the von Neumann and topological entropies of the
toric code as a function of system size and temperature. We do so for
systems with infinite energy scale separation between magnetic and
electric excitations, so that the magnetic closed loop structure is
fully preserved while the electric loop structure is tampered with by
thermally excited electric charges. 
We find that the entanglement entropy is a singular function of
temperature and system size, and that the limit of zero temperature
and the limit of infinite system size do not commute. The two orders 
of limit differ by a term that does not depend on the size of the boundary
between the partitions of the system, 
but instead depends on the topology of the bipartition. 
{}From the entanglement entropy we obtain the topological entropy, which
is shown to drop to half its zero-temperature value for any infinitesimal
temperature in the thermodynamic limit, and remains constant as the
temperature is further increased. 
Such discontinuous behavior is replaced by a smooth decreasing function 
in finite-size systems. 
If the separation of energy scales in the system is large but finite, 
we argue that our results hold at small enough temperature and finite 
system size, and a second drop in the topological entropy should occur as
the temperature is raised so as to disrupt the magnetic loop structure 
by allowing the appearance of free magnetic charges. 
We discuss the scaling of these
entropies as a function of system size, and how the quantum topological 
entropy is shaved off in this two-step process as a function of temperature
and system size.
We interpret our results as an indication that the underlying magnetic and 
electric closed loop structures contribute equally to the topological 
entropy (and therefore to the topological order) in the system. 
Since each loop structure \emph{per se} is a classical object, we 
interpret the quantum topological order in our system as arising from the 
ability of the two structures to be superimposed and appear simultaneously.
\end{abstract}

\maketitle
%
%

\section{\label{sec: intro}
Introduction
        } 

Some strongly correlated quantum systems have rather rich spectral
properties, such as ground state degeneracies that are not related to
symmetries, but instead to topology.~\cite{Haldane1985,Wen1990} 
Such systems are said to be topologically ordered,~\cite{topo refs} 
and they can have 
excitations with fractionalized quantum numbers,~\cite{Arovas1984} 
as in the case of the fractional quantum Hall states. 
There have been proposals to utilize topologically ordered
states for fault tolerant quantum computation, exploiting the
resilience of these systems to decoherence by local perturbations or
disturbances by the environment.

Levin and Wen~\cite{Levin2006}, and Kitaev and
Preskill~\cite{Kitaev2006} recently proposed that a characteristic
signature of topological order can be found in a subleading correction
of the Von Neumann (entanglement) entropy in systems prepared in (one
of) its ground state(s). This topological correction to the
entanglement entropy was indeed confirmed by exact calculations in
discrete models exhibiting topological order, as well as in continuum
systems such as fermionic Laughlin states.~\cite{Haque2007} The notion
of topological entropy provides a ``non-local order parameter'' for
topologically ordered systems. Hereafter, we refer to topological
order as characteristically identified by such non-vanishing
topological entropy.

Although quantum topological order was introduced as a pure
zero-temperature concept, is was recently shown~\cite{Castelnovo2006}
that a closely related behavior can be observed also in mixed state
density matrices that describe classical systems in the presence of
hard constraints.
These findings show that topological order can survive thermal mixing
under certain conditions, e.g., in hard constrained systems.
Moreover, any possible experimental observation of quantum topological
order must take into account the fact that the $T=0$ limit is only an
idealization and temperature, albeit small, is a perturbation that
cannot be neglected. This is particularly relevant, for example, if one
is interested in a practical application of topological order towards
quantum computing, which will always be done at finite
temperature. 
It is therefore interesting to study the behavior of topologically
ordered systems as the temperature is gradually raised from zero, in
search of a unified picture of topological order encompassing both the
quantum zero-temperature limit and the classical hard-constrained
limit.

In this paper, we investigate the fate of quantum topological order in
the two-dimensional toric code on the square lattice in thermal equilibrium 
with a bath at finite temperature. 
In particular, we do so by studying the entanglement and topological
entropies of the system, which we compute exactly. 

We start from the zero-temperature limit of the model, which has been
thoroughly studied in Ref.~\onlinecite{Kitaev2003}. In this limit, the
ground state (GS) of the system can be mapped onto two loop 
structures~\cite{Hamma2006} 
each of which, we argue, is responsible for half of the topological
contribution to the von Neumann entropy (i.e., half of the topological
entropy of the system).  As temperature is raised from zero, thermal
equilibration disrupts (breaks) the loop structure and it is expected
to destroy topological order.  With an exact calculation in the limit
where one of the two loop structures is fully preserved while the other is 
allowed to thermalize,~\cite{Trebst2007} we show that the topological entropy
gradually decreases as a function of temperature, for fixed and finite
system size, from its zero-temperature value down to precisely half of
that value.  In particular, the temperature dependence of the
topological entropy can be shown to appear always through the product
$K^{\ }_{A}(T) \,N$, where $K^{\ }_{A}(T)$ is a monotonic function
of temperature with $K^{\ }_{A}(0) = 0$ and $K^{\ }_{A}(\infty) =
\infty$, and $N$ is an extensive quantity that scales linearly with
the number of degrees of freedom in the system.  Therefore, the
thermodynamic limit $N \to \infty$ and the $T \to 0$ limit \emph{do
not commute}, and if the former is taken first, the topological
entropy becomes a singular function at $T=0$, and it equals one half
of its zero-temperature value for any $T \neq 0$.  In other words, in
the thermodynamic limit any infinitesimal temperature is able to fully
disrupt any loop structure for which we allow thermalization, and the 
contribution from this structure to the topological entropy is completely 
lost (irrespective of the presence of a finite energy gap). 
On the other hand, 
finite size systems can retain a statistical contribution to the 
topological entropy (in the sense that its value varies continuously with
temperature) originating from a thermalized underlying loop structure.

\begin{figure}[ht]
\includegraphics[width=0.9\columnwidth]{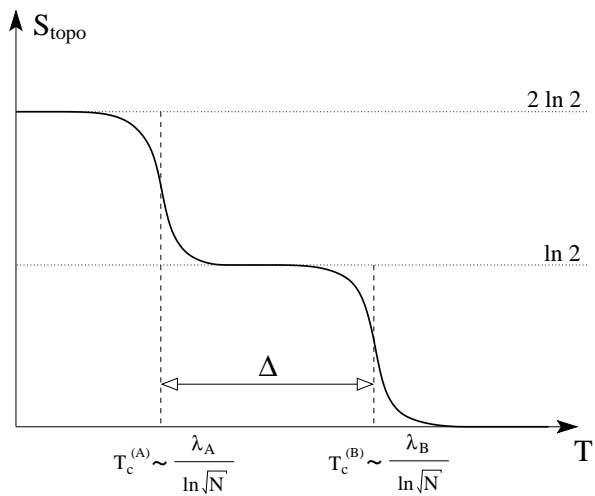}
\caption{
\label{fig: S_topo vs T full}
Qualitative behavior of the topological entropy as a function of
temperature $T$ and number of degrees of freedom $2N$, 
for the generic case where the
two coupling constants in the model are well separated, i.e.,
$\lambda^{\ }_{A} \ll \lambda^{\ }_{B}$. The exact shape of the first
crossover is shown in Fig.~\ref{fig: KN collapse}}
\end{figure}

{}From our results, we then infer the behavior of the finite-temperature 
topological entropy in the generic case, as illustrated in 
Fig.~\ref{fig: S_topo vs T full}. 
For finite size systems, we expect to observe two continuous decays of
the topological entropy, due to the gradual disruption of each of its
two loop contributions. Each drop occurs when the number of
corresponding defects $\sim N\;e^{-2\lambda^{\ }_{A,B}/T}_{\ }$ reaches 
a value of order one, where $2N$ is the number of degrees of freedom in the 
system, and $\lambda^{\ }_{A}$ and $\lambda^{\ }_{B}$ are the two coupling 
constants in the model, associated with one loop structure each. 
The separation between the two decays (the quantity
$\Delta$ in the figure) is therefore proportional to the difference
between the two coupling constants $\lambda^{\ }_{A}$ and 
$\lambda^{\ }_{B}$ in the Hamiltonian. 
Once again, if the thermodynamic limit is taken first, both decays
collapse into a singular behavior where the topological entropy
vanishes everywhere except for $T=0$, where its value depends on the 
order between the thermodynamic and zero-temperature limit. 
Notice also that, although the crossover temperature 
$T^{(A,B)}_{\rm cross} \sim \lambda^{\ }_{A,B} / \ln\sqrt{N}$ goes to 
zero in the limit of $N \to \infty$, it does so only in a logarithmic 
fashion. 

The case of classical topological order is recovered in the present study 
when thermal fluctuations are allowed to completely break one of the loop 
structures while the other is strictly preserved. 
Indeed, this can be accomplished by imposing appropriate (local) hard 
constraints on the classical analog of the toric code.~\cite{Castelnovo2006} 
Our results illustrate how both the concept of quantum 
topological order and of classical topological order are equally fragile, 
as they truly exist only in the zero-temperature / hard-constraint 
limit. Their effects however can extend well into the finite-temperature 
/ soft-constraint realm -- as our calculations show -- so long as the 
size of the system is finite.

Our results suggest a simple pictorial interpretation of quantum
topological order, at least for systems where there is an easy
identification of loop structures as in the case here studied. The
picture is that (i) the two loop structures contribute equally and
independently to the topological order at zero temperature; (ii) each
loop structure \emph{per se} is a classical (non-local) object
carrying a contribution of $\ln D$ to the topological entropy 
($D=2$ being the so-called quantum dimension of the system); and
(iii) the quantum nature of the zero-temperature system resides in the
fact that two independent loop structures can be superimposed 
(therefore leading to an overall topological entropy equal to 
$2\ln D = \ln D^{2}_{\ }$). In this
sense, our results lead to an interpretation of quantum topological
order, at least for systems with simple loop structures, as the
quantum mechanical version of a classical topological order (given by
each individual loop structure).

We also investigate the von Neumann (entanglement) entropy 
$S^{\ }_{\textrm{VN}}$ as a function of temperature and system size. 
For instance, we show that, 
given any bipartition $(\A,\B)$ of the whole system 
$\mathcal{S} = \A \cup \B$, the quantity
\bea
\Delta S 
&=& 
\lim_{T \to 0, L \to \infty} S^{\A}_{\textrm{VN}}(T)
- 
\lim_{L \to \infty, T \to 0} S^{\A}_{\textrm{VN}}(T) 
\nonumber \\ 
&=& 
(m^{\ }_{\B}-1)\ln 2, 
\eea
where $S^{\A}_{\textrm{VN}}(T)$ is the entropy of partition $\A$
after tracing out partition $\B$, $m^{\ }_{\B}$ is the number of
disconnected regions in $\B$, and $L = \sqrt{N}$ is the linear size of the 
system. 
{}From this result, we learn that the
topological contribution to the entanglement entropy can be filtered
out directly from a single bipartition, provided $m^{\ }_{\B} > 1$, as
opposed to the constructions in
Refs.~\onlinecite{Levin2006,Kitaev2006} that require a linear
combination over multiple bipartitions.

We also show that, as soon as the temperature is different from zero,
the von Neumann entropy is no longer symmetric upon exchange of
subsystem $\A$ and subsystem $\B$, and it acquires a term that is
extensive in the number of degrees of freedom that have not been
traced out (see Eq.~(\ref{eq: S_VN(T) for large L})). Symmetry and
dependence only on the boundary degrees of freedom, at least in the 
thermodynamic limit, can be recovered if 
one considers instead the mutual information
\bea
I^{\ }_{\A\B}(T) 
&=& 
\frac{1}{2}
  \left[\vphantom{\sum} 
    S^{\A}_{\textrm{VN}}(T) 
    +
    S^{\B}_{\textrm{VN}}(T)
    -
    S^{\A \cup \B}_{\textrm{VN}}(T)
  \right]
\;, 
\eea
as we explicitly show in this paper. Notice that 
$
I^{\ }_{\A\B}(0) 
\equiv 
S^{\A}_{\textrm{VN}}(0) 
= 
S^{\B}_{\textrm{VN}}(0)
$. 
Once again, the mutual information exhibits a singular behavior at zero 
temperature since the thermodynamic limit and the zero-temperature limit 
do not commute. 
We find that the explicit topological contribution to $I^{\ }_{\A\B}(T)$ 
in the thermodynamic limit is
$
-\frac{1}{2} 
\left( 
  m^{\ }_{\A} + m^{\ }_{\B} - 1 
\right) 
  \ln 2
$, where $ m^{\ }_{\A}$ $(m^{\ }_{\B})$ is the number of  disconnected 
components of partition $\A$ $(\B)$.

Although we consider here a very specific model, we believe that our
results are of relevance to a broader context, at least at a
qualitative level. For example, it would be interesting to investigate
the specific behavior of systems where the underlying structures
responsible for the presence of topological order are no longer
identical to each other. This is the case of the three-dimensional
extension of the toric code, where a closed loop structure becomes
dual to a closed membrane structure.~\cite{Wen-comment}

The paper is organized as follows. In Section~\ref{sec: the model} we 
present the model and discuss its characteristic features and properties. 
In Section~\ref{sec: S_VN} we compute the von Neumann entropy of the 
system as a function of temperature and system size, in the limit of 
one of the coupling constants going to infinity. 
We then obtain the exact expression for the topological entropy in 
Sec.~\ref{sec: topo entropy}, and we illustrate its behavior with 
numerical results. 
Finally, we discuss the implications of our results for the system with 
finite coupling constants and we infer the full temperature and system size 
dependence of the topological entropy in 
Sec.~\ref{sec: full temperature range}. 
Conclusions are drawn in Sec.~\ref{sec: conclusions}. 
%
%

\section{\label{sec: the model} 
The finite-temperature toric code
        }
The zero-temperature limit of the model considered here was studied by 
Kitaev in Ref.~\onlinecite{Kitaev2003}. It can be represented 
by $2N$ spin-$1/2$ degrees of freedom on the bonds of an $L \times L$ square 
lattice, $N = L^{2}_{\ }$, with periodic boundary conditions 
(toric geometry). 
The system is endowed with a Hamiltonian that can be written in terms of 
star and plaquette operators as 
\beq
H 
= 
-\lambda^{\ }_{B} \sum^{\ }_{\textrm{plaquettes}\:p} B^{\ }_{p} 
-\lambda^{\ }_{A} \sum^{\ }_{\textrm{stars}\:s} A^{\ }_{s}, 
\label{eq: Kitaev Hamiltonian}
\eeq
where $\lambda^{\ }_{A}$ and $\lambda^{\ }_{B}$ are two positive coupling 
constants, 
$B^{\ }_{p} = \prod^{\ }_{i \in p} \sigma^{\textrm{z}}_{i}$, 
and 
$A^{\ }_{s} = \prod^{\ }_{j \in s} \sigma^{\textrm{x}}_{j}$, 
with $i$ labeling all four edges of plaquette $p$ and 
$j$ labeling all four bonds meeting at vertex $s$ of the square lattice. 
Notice that the Hamiltonian, all the $B^{\ }_{p}$ operators and 
all the $A^{\ }_{s}$ commute with each other, and one can diagonalize them 
simultaneously. 
Given that there are $N-1$ independent plaquette operators and $N-1$ 
independent star operators 
($\prod^{N}_{p=1} B^{\ }_{p} = \openone = \prod^{N}_{s=1} A^{\ }_{s}$), the 
eigenvectors with fixed $B^{\ }_{p}$ and $A^{\ }_{s}$ quantum numbers form 
a $2^{2}_{\ }$-dimensional space. 
Furthermore, one can show that the GS $4$-fold degeneracy has a topological 
nature that can be split only by the action of non-local (system spanning) 
operators. 
The ground state wavefunctions of this model are known 
exactly,~\cite{Kitaev2003} and can be written in the 
$\sigma^{\textrm{z}}_{\ }$ basis as 
\beq
\vert \Psi^{\ }_{0} \rangle 
= 
\frac{1}{\vert G \vert^{1/2}_{\ }} 
  \sum^{\ }_{g \in G} g \vert 0 \rangle, 
\label{eq: Kitaev GS}
\eeq
where $G$ is the Abelian group generated by all star operators 
$\{ A^{\ }_{s} \}^{N}_{s=1}$, modulo the fact that 
$\prod^{N}_{s=1} A^{\ }_{s} = \openone$, 
$\vert G \vert = 2^{N-1}_{\ }$ is the dimension of $G$, 
and 
$
\vert 0 \rangle 
= 
\vert \sigma^{\textrm{z}}_{1} \,\ldots\, \sigma^{\textrm{z}}_{n} \rangle
$
is any given state that satisfies the condition 
$B^{\ }_{p}\vert 0 \rangle = \vert 0 \rangle$, $\forall\,p$. 
There are four inequivalent choices for $\vert 0 \rangle$, corresponding 
to the four different topological sectors of the model. 
The choice of sector is immaterial to the results presented hereafter, 
since they all have the same entanglement,~\cite{Hamma2005} 
and we will set $\vert 0 \rangle = \vert ++\ldots+ \rangle$ for convenience 
throughout the rest of the paper. 

\begin{figure}[ht]
\vspace{0.2 cm}
\includegraphics[width=0.98\columnwidth]{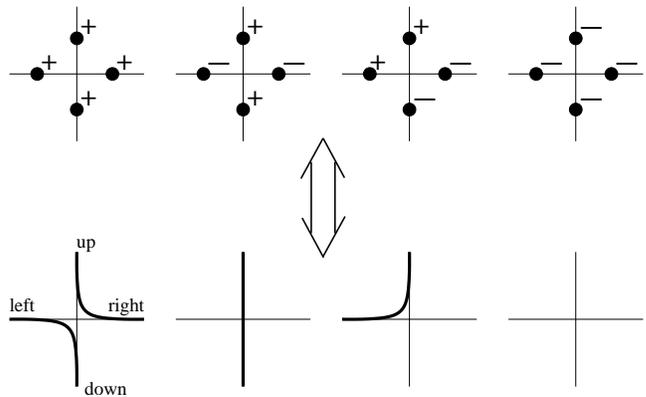}
\caption{
\label{fig: spin to loop}
Illustration of the spin-loop correspondence discussed in the text. 
All vertices with two positive and two negative $\sigma^{\textrm{x}}_{\ }$ 
components on the adjacent links can be obtained via appropriate rotations 
of the ones shown in the figure. 
}
\end{figure}

Notice that the system is symmetric upon exchange of
$\sigma^{\textrm{x}}_{\ }$ with $\sigma^{\textrm{z}}_{\ }$ components
and of stars with plaquettes on the lattice.  In the
$\sigma^{\textrm{x}}_{\ }$-basis, each site must have $0$, $2$ or $4$
spins with a negative $\sigma^{\textrm{x}}_{\ }$ component on the
adjacent bonds. If we were to remove all the bonds with a negative
$\sigma^{\textrm{x}}_{\ }$ component, we would obtain a configuration
of closed loops on the square lattice, where loops are allowed to
cross but do not overlap.  Once a convention is established on how to
interpret sites entirely surrounded by spins with positive
$\sigma^{\textrm{x}}_{\ }$ component (e.g., as two different loop
parts touching at the corner, say the up-right and down-left loops),
then one can establish a one-to-one correspondence between all basis
states and all loop configurations on the square lattice where loops
cannot overlap and can at most touch at a corner in an up-right,
down-left fashion (see Fig.~\ref{fig: spin to loop}).
The same is true for the $\sigma^{\textrm{z}}_{\ }$-basis, but the loops 
now live on the dual lattice given by the centers of the plaquettes of 
the original lattice. 
This description of the GS of the toric code in terms of loop degrees 
of freedom gives a qualitative picture of the origin of the non-local 
behavior of the system from which the presence of topological order stems. 
In particular, given that the two simultaneous loop descriptions can be 
mapped one onto the other upon exchanging $\sigma^{\textrm{x}}_{\ }$ with 
$\sigma^{\textrm{z}}_{\ }$ as well as the square lattice with its dual, 
it is tempting to speculate that they contribute equally to the topological 
order present in the system, and each loop structure is responsible for 
precisely half of the topological entropy. 
Our exact calculations show that this naive picture is indeed correct: 
if either of the loop structures is exactly preserved while the other is 
destroyed, e.g., via coupling to a thermal bath, the topological entropy 
of the system lowers to half of its original value. 

Based on the $\sigma^{\textrm{x}}_{\ }$ and $\sigma^{\textrm{z}}_{\ }$ loop 
description, all possible perturbations to the system can be qualitatively 
divided into three different classes: 
(i) those that couple to a $\sigma^{\textrm{z}}_{\ }$-like term, and are 
able -- if sufficiently strong -- 
to disrupt the underlying $\sigma^{\textrm{x}}_{\ }$ loop structure, 
but not the $\sigma^{\textrm{z}}_{\ }$ one; 
(ii) those that couple to a $\sigma^{\textrm{x}}_{\ }$-like term, with 
precisely the opposite effect; 
and 
(iii) those that couple to a $\sigma^{\textrm{y}}_{\ }$-like term, and 
are able -- again, if the coupling constant is large enough -- 
to disrupt both loop structures, thus leading to a vanishing 
topological entropy. 
A generic coupling to a thermal bath is likely to encompass all of the 
above terms and in the thermodynamic limit the vanishing of the topological 
entropy is unavoidable. 
For finite size systems and at low enough temperature, however, 
the relative scale of the two coupling constants $\lambda^{\ }_{B}$ and 
$\lambda^{\ }_{A}$ plays a crucial role in determining how effective each 
of the above terms is with respect to the others. 
In this paper we consider the case when the two energy scales are well 
separated, namely $\lambda^{\ }_{B} \gg \lambda^{\ }_{A}$, and we discuss 
qualitatively the behavior of the system as the separation becomes weaker 
and vanishes. 
A large separation between the two energy scales is indeed expected if we 
notice that the toric code is a lattice realization of a 
$\mathbb{Z}^{\ }_{2}$ gauge theory, where the two coupling constants 
$\lambda^{\ }_{A}$ and $\lambda^{\ }_{B}$ relate directly to the chemical 
potential of the electric and magnetic monopoles.~\cite{gauge_refs} 
On the ground of a large separation between the 
two energy scales in the Hamiltonian, three distinct temperature 
regimes can be outlined: 
\begin{itemize}
\item[(a)] $ T\ll \lambda^{\ }_{A}/\ln\sqrt{N}$, when all thermal
excitations have a small Boltzmann weight and for finite size systems
at finite time scales the topological entropy effectively retains its
zero temperature value because of the scarcity of defects that can
disrupt the loop structure;
\item[(b)] $\lambda^{\ }_{A}/\ln\sqrt{N} < T \ll \lambda^{\
}_{B}/\ln\sqrt{N}$, when thermal excitations of the
$\sigma^{\textrm{z}}_{\ }$ type can disrupt the
$\sigma^{\textrm{x}}_{\ }$ loops structure, while the
$\sigma^{\textrm{x}}_{\ }$-like excitations are rather unlikely
to occur and they can be effectively neglected;
and 
\item[(c)]
$\lambda^{\ }_{B}/\ln\sqrt{N} < T$, when the appearance of all the three types of 
thermal excitations leads to the complete disruption of the topological 
contribution to the entanglement entropy.
%
\end{itemize}
(Notice that the opposite case, where 
$\lambda^{\ }_{A} \gg \lambda^{\ }_{B}$, leads to equivalent results 
based on the symmetry of the model.) 
The temperature range considered in this paper corresponds to regimes~(a) 
and~(b), where the $\sigma^{\textrm{z}}_{\ }$ loop structure is effectively 
preserved for sufficiently small system sizes and time scales. 
We will then discuss how our results can be used to infer the behavior of 
the topological entropy across the whole temperature range, illustrated 
in Fig.~\ref{fig: S_topo vs T full}. 

Basically, one can define a temperature dependent defect separation
length scales $\xi^{\ }_{A,B}\sim e^{\lambda^{\ }_{A,B} /T}$, so that
as long as the sample size is below these corresponding scales, the
system is free from the associated type of defects. Similarly, one can
define temperature-dependent time scales for defects to appear. The
toric code is {\it fragile} in the sense that ${\cal O}(1)$ defects
destroy its topological order, so that for practical considerations
not only the temperature must be small compared to a gap, but the
system size and the time scales must not be too large as well.

We now consider the simplification where the finite system length
$\sqrt{N}\ll \xi^{}_{B}$, so we can neglect defects in the
$\sigma^{\textrm{z}}_{\ }$ loop structure. Forbidding any defects in
the $\sigma^{\textrm{z}}_{\ }$ loop structure is equivalent to
neglecting all thermal processes that violate the
constraint 
$\prod^{\ }_{i \in p} \sigma^{\textrm{z}}_{i} = +1$, 
$\forall\,p$. 
Therefore, the Hilbert space in the regime of interest and within the 
chosen topological sector (recall that 
$\vert 0 \rangle = \vert + + \ldots + \rangle$) is 
given by 
$\{ g \vert 0 \rangle \;\vert\; g\in G\}$. 
The equilibrium properties of the system are then captured 
by the finite-temperature density matrix 
\bea
\rho(T) 
&=& 
\frac{1}{Z} \: e^{-\beta \hat H} 
\nonumber \\ 
&=& 
\frac{\sum_{g,g^{\prime}_{\ } \in G} 
      \langle 0 \vert 
        g e^{-\beta H}_{\ } gg^{\prime}_{\ } 
      \vert 0 \rangle
      \;
      g \vert 0 \rangle \langle 0 \vert gg^{\prime}_{\ } 
     }
     {\sum_{g \in G} 
      \langle 0 \vert 
        g e^{-\beta H} g 
      \vert 0 \rangle
     }, 
\label{eq: rho(T) def}
\eea
where we used the group property to write a generic element 
$g^{\prime\prime}_{\ } \in G$ as 
$g^{\prime\prime}_{\ } = gg^{\prime}_{\ }$, 
$\exists ! \, g^{\prime}_{\ } \in G$ given $g \in G$. 
Recall that all group elements, as well as their composition, are defined 
as products of star operators 
modulo the identity $\prod^{N}_{s=1} A^{\ }_{s} = \openone$. 

For the model under consideration, it is convenient to rewrite the 
Hamiltonian~(\ref{eq: Kitaev Hamiltonian}) as 
\bea
H 
&=& 
-\lambda^{\ }_{B} P 
-\lambda^{\ }_{A} S 
\nonumber \\
P &=& \sum^{\ }_{\textrm{plaquettes}\:p} B^{\ }_{p} 
\nonumber \\
S &=& \sum^{\ }_{\textrm{stars}\:s} A^{\ }_{s}. 
\eea
Notice that $P g \vert 0 \rangle = N g \vert 0 \rangle$, 
$\forall\,g \in G$, and therefore 
\bea
\langle 0 \vert g e^{-\beta H} gg^{\prime}_{\ } \vert 0 \rangle 
&=&
e^{\beta \lambda^{\ }_{B} N}_{\ }
\langle 0 \vert 
  g e^{\beta\lambda^{\ }_{A} S}_{\ } gg^{\prime}_{\ } 
\vert 0 \rangle
\nonumber \\ 
&=&
e^{\beta \lambda^{\ }_{B} N}_{\ }
\langle 0 \vert 
  e^{\beta\lambda^{\ }_{A} S}_{\ } g^{\prime}_{\ } 
\vert 0 \rangle
\eea
where we used the fact that any $g$ commutes with $S$ by construction. 

Now, recall the definition of a group element $g \in G$, which can be 
symbolically represented by the notation 
$g \equiv \prod^{\ }_{s \in g} A^{\ }_{s}$ 
(modulo the identity $\prod^{N}_{s=1} A^{\ }_{s} = \openone$, i.e., 
$g = \prod^{\ }_{s \in g} A^{\ }_{s} = \prod^{\ }_{s \notin g} A^{\ }_{s}$). 
Given the expansion 
\beq
e^{\beta\lambda^{\ }_{A} S} 
=
\prod_{\textrm{stars}\, s} 
  \left[
    \cosh\beta\lambda^{\ }_{A} 
    +
    \sinh\beta\lambda^{\ }_{A}\: A^{\ }_{s} 
  \right], 
\eeq
which follows from the definition $S = \sum^{N}_{s=1} A^{\ }_{s}$ and from the 
fact that $A^{2}_{s} \equiv \openone$, one obtains 
\bea
&& 
\langle 0 \vert 
  e^{\beta\lambda^{\ }_{A} S}_{\ } g^{\prime}_{\ } 
\vert 0 \rangle 
= 
\nonumber \\ 
&& 
\qquad
= 
\langle 0 \vert 
  \prod_{s}
    \left[
      \cosh\beta\lambda^{\ }_{A} 
      + 
      \sinh\beta\lambda^{\ }_{A}\: A^{\ }_{s} 
    \right] 
  \prod_{s^{\prime}_{\ }\in g^{\prime}_{\ }} A^{\ }_{s^{\prime}_{\ }} 
\vert 0 \rangle 
\nonumber \\ 
&& 
\qquad
= 
\left( 
  \cosh\beta\lambda^{\ }_{A}
\right)^N 
\; 
\left( 
  \tanh\beta\lambda^{\ }_{A} 
\right)^{n(g^{\prime}_{\ })} 
\nonumber \\ 
&& 
\qquad\;\;\;\;
+ 
\left( 
  \cosh\beta\lambda^{\ }_{A} 
\right)^N 
\; 
\left( 
  \tanh\beta\lambda^{\ }_{A}
\right)^{N-n(g^{\prime}_{\ })}, 
\label{eq: exp S}
\eea
where $N$ is the total number of stars in the system, and 
$n(g^{\prime}_{\ })$ is the number of flipped stars in $g^{\prime}_{\ }$. 
Notice that the ambiguity in the definition of 
$
g 
= 
\prod^{\ }_{s \in g} A^{\ }_{s} 
\equiv 
\prod^{\ }_{s \notin g} A^{\ }_{s}
$ 
-- namely the fact that if $g^{\prime}_{\ }$ is given by the product of a 
set of $A^{\ }_{s}$, it is also given by the product of all other 
$A^{\ }_{s}$ but for those in the set -- does not affect the equation above. 
In fact, this ambiguity amounts to the mapping 
$
n(g^{\prime}_{\ })
\leftrightarrow
N-n(g^{\prime}_{\ })
$. 
Similarly, 
\bea
Z 
&=& 
\sum^{\ }_{g \in G} 
\langle 0 \vert 
  g e^{-\beta H} g 
\vert 0 \rangle 
\nonumber \\ 
&=& 
\vert G \vert \, 
e^{\beta\lambda^{\ }_{B} N}_{\ } 
\langle 0 \vert 
  \prod_{\textrm{stars}\, s}
    \left[\vphantom{\sum} 
      \cosh\beta\lambda^{\ }_{A}
      +
      \sinh\beta\lambda^{\ }_{A}\: A^{\ }_{s}
    \right] 
\vert 0 \rangle 
\nonumber \\ 
&=& 
\vert G \vert \, 
e^{\beta\lambda^{\ }_{B} N}_{\ } 
(\cosh\beta\lambda^{\ }_{A})^N 
\left[\vphantom{\sum} 
  1 
  + 
  (\tanh\beta\lambda^{\ }_{A})^N
\right]. 
\label{eq: Z in rho}
\eea

Substituting Eq.~(\ref{eq: exp S}) and Eq.~(\ref{eq: Z in rho}) into 
Eq.~(\ref{eq: rho(T) def}) after relabeling 
$K^{\ }_{A} = -\ln [\tanh(\beta\lambda^{\ }_{A})]$ 
gives 
\beq
\rho(T) 
= 
\sum_{g,g^{\prime}_{\ } \in G} 
\frac{1}{\vert G \vert}
\frac{ 
  \left[ 
    e^{-K^{\ }_{A} n(g^{\prime}_{\ })}_{\ } 
    + 
    e^{-K^{\ }_{A} \left( N - n(g^{\prime}_{\ }) \right)}_{\ } 
  \right] 
     } 
     {
  \left[ 
    1 
    + 
    e^{-K^{\ }_{A} N} 
  \right] 
     } 
\;
g \vert 0 \rangle \langle 0 \vert gg^{\prime}_{\ }.
\label{eq: rho(T)}
\eeq
In the limit of $T \to 0$ ($\beta \to \infty$), $K^{\ }_{A} \to 0^{+}_{\ }$, 
all $g^{\prime}_{\ }$ are equally weighed, and one recovers the density 
matrix of the zero-temperature Kitaev model. 
In the limit $T \to \infty$ ($\beta \to 0$), $K^{\ }_{A} \to \infty$, 
all $g^{\prime}_{\ }$ are exponentially suppressed except for 
$g^{\prime}_{\ } = \openone$, and 
one recovers the mixed-state density matrix of the topologically ordered 
classical system discussed in Ref.~\onlinecite{Castelnovo2006}. 
%
%

\section{\label{sec: S_VN}
The Von Neumann entropy
        }
Let us consider a generic bipartition of the system $\mathcal{S}$ into 
subsystem $\A$ and subsystem $\B$ ($\mathcal{S} = \A \cup \B$). 
Let us also define $\Sigma^{\ }_{\A}$ ($\Sigma^{\ }_{\B}$) to be the number 
of star operators $A^{\ }_{s}$ that act solely on spins in $\A$ ($\B$), 
and $\Sigma^{\ }_{\A\B}$ as the number of star operators acting 
simultaneously on both subsystems. 
Clearly these quantities satisfy the relationship 
$\Sigma^{\ }_{\A} + \Sigma^{\ }_{\B} + \Sigma^{\ }_{\A\B} = N$. 
Whenever a partition is made up of multiple connected components, 
e.g., $\A = \A^{\ }_{1} \cup \,\ldots\, \cup \A^{\ }_{m^{\ }_{\A}}$ with 
$\A^{\ }_{i} \cap \A^{\ }_{j} = \emptyset$ and $\A^{\ }_{i}$ connected, 
$\forall\,i,j$, 
let us denote with $\Sigma^{\ }_{\A^{\ }_{i}}$ the number of star operators 
acting solely on $\A^{\ }_{i}$ 
($\Sigma^{\ }_{\A} = \sum^{\ }_{i} \Sigma^{\ }_{\A^{\ }_{i}}$). 
[Since in the following we will consider only the case where either $\A$ 
or $\B$ have multiple connected components, but not both at the same time, 
$\Sigma^{\ }_{\A\B^{\ }_{i}}$ will be used unambiguously to denote the 
number of star operators acting simultaneously on $\A^{\ }_{i}$ and on 
$\B$ or on $\B^{\ }_{i}$ and on $\A$, according to the specific case 
($\Sigma^{\ }_{\A\B} = \sum^{\ }_{i} \Sigma^{\ }_{\A\B^{\ }_{i}}$).] 

The von Neumann (entanglement) entropy $S^{\ }_{\textrm{VN}}$ of a 
bipartition $(\A,\B)$ is given by 
\beq
S^{A}_{\textrm{VN}} 
\equiv 
-\textrm{Tr} \left[ \rho^{\ }_{A} \ln \rho^{\ }_{A} \right] 
= 
S^{B}_{\textrm{VN}}, 
\label{eq: S_VN}
\eeq
where $\rho^{\ }_{A} = \textrm{Tr}^{\ }_{\B} (\rho)$ is the reduced density 
matrix obtained from the full density matrix $\rho$ by tracing out the 
degrees of freedom of subsystem $B$, and the last equality holds whenever 
the full density matrix $\rho$ is a pure-state density matrix. 

In order to compute the von Neumann entropy~(\ref{eq: S_VN}) from the 
finite-temperature density matrix~(\ref{eq: rho(T)}), we first obtain the 
reduced density matrix of the system using the same approach of 
Ref.~\onlinecite{Hamma2005}, 
\bea
\rho^{\ }_{\A}(T) 
&=& 
\sum_{g,g^{\prime}_{\ } \in G} 
\frac{1}{\vert G \vert}
\frac{ 
  \left[ 
    e^{-K^{\ }_{A} n(g^{\prime}_{\ })}_{\ } 
    + 
    e^{-K^{\ }_{A} \left( N - n(g^{\prime}_{\ }) \right)}_{\ } 
  \right] 
     } 
     {
  \left[ 
    1 
    + 
    e^{-K^{\ }_{A} N} 
  \right] 
     }
\times
\nonumber \\ 
&& 
\qquad\qquad
g^{\ }_{\A} \vert 0^{\ }_{\A} \rangle 
  \langle 0^{\ }_{\A} \vert g^{\ }_{\A} g^{\prime}_{\A} 
\;
\langle 0^{\ }_{\B} \vert 
  g^{\ }_{\B} g^{\prime}_{B} g^{\ }_{\B} 
\vert 0^{\ }_{\B} \rangle 
\nonumber \\ 
&=& 
\sum_{g \in G, g^{\prime}_{\ } \in G^{\ }_{\A}} 
\frac{1}{|G|}
\frac{ 
  \left[ 
    e^{-K^{\ }_{A} n(g^{\prime}_{\ })}_{\ } 
    + 
    e^{-K^{\ }_{A} \left( N - n(g^{\prime}_{\ }) \right)}_{\ } 
  \right] 
     } 
     {
  \left[ 
    1 
    + 
    e^{-K^{\ }_{A} N} 
  \right] 
     }
\times
\nonumber \\ 
&& 
\qquad\qquad
g^{\ }_{\A} \vert 0^{\ }_{\A} \rangle 
  \langle 0^{\ }_{\A} \vert g^{\ }_{\A} g^{\prime}_{\A} 
\nonumber \\ 
&\equiv& 
\frac{1}{|G|} 
\sum^{\ }_{g \in G, g^{\prime}_{\ } \in G^{\ }_{\A}} 
\eta^{\ }_{T}(g^{\prime}_{\ }) 
\; 
g^{\ }_{\A} \vert 0^{\ }_{\A} \rangle 
  \langle 0^{\ }_{\A} \vert g^{\ }_{\A} g^{\prime}_{\A}, 
\label{eq: rho_A(T)}
\eea
where we used the generic tensor decomposition 
$
\vert 0 \rangle 
= 
\vert 0^{\ }_{\A} \rangle 
\otimes 
\vert 0^{\ }_{\B} \rangle
$, 
$g = g^{\ }_{\A} \otimes g^{\ }_{\B}$, 
and the fact that 
$
\langle 0^{\ }_{\B} \vert 
  g^{\ }_{\B} g^{\prime}_{\B} g^{\ }_{\B} 
\vert 0^{\ }_{\B} \rangle
= 
1
$ 
if $g^{\prime}_{\B} = \openone^{\ }_{\B}$ and zero otherwise. 
The latter follows immediately from the fact that the group $G$ is Abelian 
and that $A^{2}_{s} = \openone$, $\forall\, s$, and therefore 
$g^{2}_{\B} = \openone^{\ }_{\B}$ for any choice of $\B$. 
We also denoted by 
$
G^{\ }_{\A} 
= 
\{ 
g \in G 
\;\vert\; 
g^{\ }_{\B} = \openone^{\ }_{\B}
\}
$ 
the subgroup of $G$ given by all operations $g$ that act trivially on $\B$ 
(similarly for $G^{\ }_{\B}$ in the following). 
For convenience of notation we defined 
\beq
\eta^{\ }_{T}(g^{\prime}_{\ }) 
= 
\frac{
\left[ 
  e^{-K^{\ }_{A} n(g^{\prime}_{\ })}_{\ } 
  + 
  e^{-K^{\ }_{A} \left( N - n(g^{\prime}_{\ }) \right)}_{\ } 
\right] 
}
{
\left[ 
  1 
  + 
  e^{-K^{\ }_{A} N} 
\right]
}. 
\label{eq: eta}
\eeq

Notice that a star operator $A^{\ }_{s}$ can either act solely on spins in 
partition $\A$ (represented in the following by the notation $s \in \A$), 
solely on spins in partition $\B$ ($s \in \B$), or simultaneously on spins 
belonging to $\A$ and $\B$ (which we will refer to as 
\emph{boundary star operators}, and represent by $s \in \A\B$). 
As discussed in Ref.~\onlinecite{Castelnovo2006}, a complete set of 
generators for the subgroup $G^{\ }_{\A}$ can be constructed by taking: 
(i) all star operators that act solely on $\A$, i.e., 
$\{ A^{\ }_{s} \;\vert\; s \in \A\}$, 
together with 
(ii) the collective operators defined as the product of all stars acting 
solely on a connected component of $\B$ times the product of all boundary 
stars of that specific component, for all the 
$m^{\ }_{\B}$ connected components of $\B$, i.e., 
$
\{ 
  \prod^{\ }_{s \in \B^{\ }_{i}} A^{\ }_{s} 
  \times 
  \prod^{\ }_{s^{\prime}_{\ } \in \A\B^{\ }_{i}} A^{\ }_{s^{\prime}_{\ }}, 
\;\forall\,\textrm{connected components}\,i
\}
$. 
Notice that not all the collective operators are new operators with respect to 
those generated by the star operators in $\A$. In fact, 
$
\prod^{\ }_{i}
\left( 
  \prod^{\ }_{s^{\prime}_{\ } \in \B^{\ }_{i}} A^{\ }_{s^{\prime}_{\ }} 
  \prod^{\ }_{s^{\prime\prime}_{\ } \in \A\B^{\ }_{i}} A^{\ }_{s^{\prime\prime}_{\ }}
\right)
\equiv 
\prod^{\ }_{s \in \A} A^{\ }_{s}
$, 
and one can show that there are precisely $m^{\ }_{\B}-1$ new, independent 
operators. 
Consequently, the cardinality of the subgroup $G^{\ }_{\A}$ is given by 
$
d^{\ }_{\A} 
\equiv 
\vert G^{\ }_{\A} \vert 
= 
2^{\Sigma^{\ }_{A}+m^{\ }_{\B}-1}_{\ }
$. 
Similarly for $G^{\ }_{\B}$, 
$
d^{\ }_{\B} 
\equiv 
\vert G^{\ }_{\B} \vert 
= 
2^{\Sigma^{\ }_{B}+m^{\ }_{\A}-1}_{\ }
$. 

To proceed with the calculation of the von Neumann entropy of the finite 
temperature system, it is useful to use the above set of generators in order 
to represent the group $G^{\ }_{\A}$ in terms of Ising spin variables 
$
\{ 
  \theta^{\ }_{s},\,\Theta^{\ }_{i} 
\}^{\textrm{connected components}\,i}_{s \in \A}
$, 
where $\theta^{\ }_{s} = -1$ ($1$) corresponds to the star operator 
$A^{\ }_{s}$ appearing (not appearing) in the decomposition of 
$g \in G^{\ }_{\A}$, and similarly $\Theta^{\ }_{i} = -1$ ($1$) corresponds 
to the collective operator 
$
\prod^{\ }_{s \in \B^{\ }_{i}} A^{\ }_{s} 
\times 
\prod^{\ }_{s^{\prime}_{\ } \in \A\B^{\ }_{i}} A^{\ }_{s^{\prime}_{\ }}
$ 
appearing (not appearing) in the same decomposition. 
Notice that the correspondence is $2$-to-$1$, since a configuration 
$\{ \theta^{\ }_{s},\,\Theta^{\ }_{i} \}^{\ }_{i,s}$ and its spin-flipped 
counterpart 
$\{ \overline{\theta}^{\ }_{s},\,\overline{\Theta}^{\ }_{i} \}^{\ }_{i,s}$, 
where $\overline{\theta}^{\ }_{s} = -\theta^{\ }_{s}$ and 
$\overline{\Theta}^{\ }_{i} = -\Theta^{\ }_{i}$, map onto the exact same 
$g \in G^{\ }_{\A}$ (which follows from the fact that one of the collective 
operators can be generated out of the others appropriately combined with the 
star operators in $\A$). 

In this representation, 
\begin{subequations}
\bea
n(g) 
&=& 
\sum^{\ }_{s} \frac{1-\theta^{\ }_{s}(g)}{2} 
+ 
\sum^{\ }_{i} 
  \left( 
    \Sigma^{\ }_{\B^{\ }_{i}} 
    + 
    \Sigma^{\ }_{\A\B^{\ }_{i}} 
  \right) 
    \frac{1-\Theta^{\ }_{i}(g)}{2} 
\nonumber \\ 
&=& 
\frac{N}{2} 
- 
\frac{1}{2} \sum^{\ }_{s} \theta^{\ }_{s}(g) 
- 
\frac{1}{2} 
  \sum^{\ }_{i}   
    \Sigma^{\ }_{\P^{\ }_{i}} \Theta^{\ }_{i}(g) 
\\ 
N-n(g) 
&=& 
\sum_s \frac{1+\theta^{\ }_{s}(g)}{2} 
+ 
\sum^{\ }_{i} 
  \left( 
    \Sigma^{\ }_{\B^{\ }_{i}} 
    + 
    \Sigma^{\ }_{\A\B^{\ }_{i}} 
  \right) 
    \frac{1+\Theta^{\ }_{i}(g)}{2} 
\nonumber \\ 
&=& 
\frac{N}{2} 
+ 
\frac{1}{2} \sum^{\ }_{s} \theta^{\ }_{s}(g) 
+ 
\frac{1}{2} 
  \sum^{\ }_{i}   
    \Sigma^{\ }_{\P^{\ }_{i}} \Theta^{\ }_{i}(g), 
\eea
\label{eqs: n(g)}
\end{subequations}
where we used the fact that 
$
\Sigma^{\ }_{\A} 
+ 
\sum^{\ }_{i} 
  \left( 
    \Sigma^{\ }_{\B^{\ }_{i}} 
    + 
    \Sigma^{\ }_{\A\B^{\ }_{i}} 
  \right) 
= 
N
$ 
and we introduced the notation 
$
\Sigma^{\ }_{\P^{\ }_{i}} 
\equiv 
\Sigma^{\ }_{\B^{\ }_{i}} 
+ 
\Sigma^{\ }_{\A\B^{\ }_{i}} 
$. 

Let us then use Eq.~(\ref{eq: rho_A(T)}) to compute the $n$-th power 
of $\rho^{\ }_{\A}(T)$: 
\bea
\rho^{n}_{\A}(T) 
&=& 
\left( \frac{1}{|G|} \right)^{n}_{\ } 
\mathop{\sum_{g^{\ }_{1} \in G}}
       ^{\ }_{g^{\prime}_{1} \in G^{\ }_{\A}} 
\ldots 
\mathop{\sum_{g^{\ }_{n} \in G}}
       ^{\ }_{g^{\prime}_{n} \in G^{\ }_{\A}} 
\left( 
\prod^{n}_{l=1} 
  \eta^{\ }_{T}(g^{\prime}_{l}) 
\right) 
\times 
\nonumber \\ 
&& 
g^{\ }_{1,\A} \vert 0^{\ }_{\A} \rangle 
  \langle 0^{\ }_{\A} \vert 
    g^{\ }_{1,\A} g^{\prime}_{1,\A} g^{\ }_{2,\A} 
  \vert 0^{\ }_{\A} \rangle 
  \langle 0^{\ }_{\A} \vert 
    g^{\ }_{2,\A} g^{\prime}_{2,\A} 
\nonumber \\ 
&& 
\ldots 
    g^{\ }_{n,\A} 
  \vert 0^{\ }_{\A} \rangle 
  \langle 0^{\ }_{\A} \vert g^{\ }_{n,\A} g^{\prime}_{n,\A}. 
\eea
Each expectation value above imposes 
$g^{\ }_{l,\A} g^{\prime}_{l,\A} g^{\ }_{l+1,\A} = \openone^{\ }_{\A}$, 
$l=1,\ldots,n-1$, 
and therefore $g^{\ }_{l} g^{\prime}_{l} g^{\ }_{l+1} \in G^{\ }_{\B}$. 
Upon relabeling $n-1$ summation variables so that 
$\tilde{g}^{\ }_{l+1} \equiv g^{\ }_{l} g^{\prime}_{l} g^{\ }_{l+1}$ for 
$l=1,\ldots,n-1$, the corresponding 
sums can then be combined with the respective inner product and they can be 
written as $\sum^{\ }_{\tilde{g}^{\ }_{l+1} \in G^{\ }_{\B}} 1 = d^{\ }_{\B}$. 
Therefore, the equation above can be simplified to 
\bea
\rho^{n}_{\A}(T) 
&=& 
\frac{d^{n-1}_{\B}}{|G|^{n}_{\ }} 
\sum_{g^{\ }_{1} \in G}
\left( 
\prod^{n}_{l=1} 
  \: 
  \sum^{\ }_{g^{\prime}_{l} \in G^{\ }_{\A}}
    \eta^{\ }_{T}(g^{\prime}_{l}) 
\right) 
\times 
\nonumber \\ 
&& 
g^{\ }_{1,\A} \vert 0^{\ }_{\A} \rangle 
  \langle 0^{\ }_{\A} \vert g^{\ }_{1,\A} 
    g^{\prime}_{1,\A} \:\ldots\: g^{\prime}_{n,\A}. 
\eea

Taking the trace of $\rho^{n}_{\A}(T)$, using the fact that 
all the $g$'s commute, and 
$\sum^{\ }_{g^{\ }_{1} \in G} 1 = \vert G \vert$, one obtains 
\bea
\textrm{Tr} \left[ \rho^{n}_{\A} \right] 
&=& 
\nonumber \\ 
&& 
\!\!\!\!\!\!\!\!\!\!\!\!\!\!\!\!\!\!\!\!\!
= 
\left( \frac{d^{\ }_{\B}}{\vert G \vert} \right)^{n-1}_{\ } 
\prod^{n}_{l=1} 
  \:
  \sum^{\ }_{g^{\prime}_{l} \in G^{\ }_{\A}}
    \eta^{\ }_{T}(g^{\prime}_{l}) 
\: 
\langle 0^{\ }_{\A} \vert 
  g^{\prime}_{1,\A} \:\ldots\: g^{\prime}_{n,\A} 
\vert 0^{\ }_{\A} \rangle 
\nonumber \\ 
&& 
\!\!\!\!\!\!\!\!\!\!\!\!\!\!\!\!\!\!\!\!\!
= 
\left( \frac{d^{\ }_{\B}}{\vert G \vert} \right)^{n-1}_{\ } 
\frac{1}{2^{n}_{\ }} 
\prod^{n}_{l=1} 
  \sum^{\textrm{constr.}}_{\{ \theta^{(l)}_{s}, 
                \Theta^{(l)}_{i} 
	     \}^{\ }_{i,s}} 
    \!\!\!\!\!\eta^{\ }_{T}(\{ \theta^{(l)}_{s},\Theta^{(l)}_{i} \}), 
\label{eq: Tr(rho^n_A(T)) -- 1}
\eea
where the factor of $1/2^{n}_{\ }$ comes from the $2$-to-$1$ 
nature of the representation of $G^{\ }_{\A}$ in terms of Ising spin 
configurations, and 
the restricted summation 
$
  \sum^{\textrm{constr.}}_{\{ 
                              \theta^{(l)}_{s}, 
                              \Theta^{(l)}_{i} 
	                  \}}
$ 
is subject to the constraint 
$
\langle 0^{\ }_{\A} \vert 
  g^{\prime}_{1,\A} \:\ldots\: g^{\prime}_{n,\A} 
\vert 0^{\ }_{\A} \rangle 
\neq 
0
$, 
which can be explicitly stated in terms of the spins $\theta^{\ }_{s}$ and 
$\Theta^{\ }_{i}$ as 
\bea
&& 
\left\{ 
\prod^{\ }_{i,s} 
  \delta(\prod^{n}_{l=1} \Theta^{(l)}_{i} - 1) 
  \delta(\prod^{n}_{l=1} \theta^{(l)}_{s} - 1) 
\right. 
\nonumber \\ 
&& 
\qquad\qquad\qquad
\left. 
+ 
\prod^{\ }_{i,s} 
  \delta(\prod^{n}_{l=1} \Theta^{(l)}_{i} + 1) 
  \delta(\prod^{n}_{l=1} \theta^{(l)}_{s} + 1) 
\right\}. 
\nonumber
\eea
Above and in the following, the short-hand notation $i,s$ in sums and 
products stands for $\textrm{connected components}\,i$ and $s \in \A$. 

Let us then substitute Eqs.~(\ref{eqs: n(g)}) into Eq.~(\ref{eq: eta}), 
\bea
\eta^{\ }_{T}(\{ \theta^{(l)}_{s},\Theta^{(l)}_{i} \}) 
&=& 
\nonumber \\ 
&& 
\!\!\!\!\!\!\!\!\!\!\!\!\!\!\!\!\!\!\!\!\!\!\!\!\!\!\!\!\!\!\!\!\!
= 
\frac{
\cosh
  \left( 
    \frac{K^{\ }_{A}}{2} \sum^{\ }_{s} \theta^{(l)}_{s} 
    + 
    \frac{K^{\ }_{A}}{2} \sum^{\ }_{i}   
      \Sigma^{\ }_{\P^{\ }_{i}} \Theta^{(l)}_{i} 
  \right)
}
{
\cosh
  \left( 
    \frac{K^{\ }_{A}}{2} N 
  \right)
} 
\nonumber \\ 
&& 
\!\!\!\!\!\!\!\!\!\!\!\!\!\!\!\!\!\!\!\!\!\!\!\!\!\!\!\!\!\!\!\!\!
= 
\frac{1}{2 \cosh\left( \frac{K^{\ }_{A}}{2} N \right)} 
\times
\nonumber \\ 
&& 
\!\!\!\!\!\!\!\!\!\!\!\!\!\!\!\!\!\!\!\!\!\!\!\!\!\!\!\!\!\!\!\!\!
\qquad
\sum^{\ }_{J = \pm 1} 
  e^{\frac{K^{\ }_{A}}{2} J \sum^{\ }_{s} \theta^{(l)}_{s} 
     + 
     \frac{K^{\ }_{A}}{2} J \sum^{\ }_{i}   
       \Sigma^{\ }_{\P^{\ }_{i}} \Theta^{(l)}_{i} 
    }_{\ } 
\nonumber \\ 
&& 
\!\!\!\!\!\!\!\!\!\!\!\!\!\!\!\!\!\!\!\!\!\!\!\!\!\!\!\!\!\!\!\!\!
= 
\frac{1}{2 \cosh\left( \frac{K^{\ }_{A}}{2} N \right)} 
\times
\nonumber \\ 
&& 
\!\!\!\!\!\!\!\!\!\!\!\!\!\!\!\!\!\!\!\!\!\!\!\!\!\!\!\!\!\!\!\!\!
\qquad
\sum^{\ }_{J = \pm 1} 
  \prod^{\ }_{s \in \A} 
    e^{\frac{K^{\ }_{A}}{2} J \theta^{(l)}_{s}}_{\ } 
  \prod^{\ }_{i} 
    e^{\frac{K^{\ }_{A}}{2} J 
       \Sigma^{\ }_{\P^{\ }_{i}} \Theta^{(l)}_{i}}_{\ }, 
\eea
and take the sum over all possible 
$\{ \theta^{(l)}_{s}, \Theta^{(l)}_{i} \}^{\ }_{i,s}$ configurations 
(without any constraint), 
\bea
&& 
\left[ 
  2 \cosh\left( \frac{K^{\ }_{A}}{2} N \right) 
\right] 
\sum^{\ }_{\{ \theta^{(l)}_{s}, \Theta^{(l)}_{i} \}^{\ }_{i,s}} 
  \eta^{\ }_{T}(\{ \theta^{(l)}_{s},\Theta^{(l)}_{i} \}) 
=
\nonumber \\ 
&& 
\qquad
= 
\sum^{\ }_{J = \pm 1} 
 \sum^{\ }_{\{ \theta^{(l)}_{s}, \Theta^{(l)}_{i} \}^{\ }_{i,s}} 
  \prod^{\ }_{s \in \A} 
    e^{\frac{K^{\ }_{A}}{2} J \theta^{(l)}_{s}}_{\ } 
  \prod^{\ }_{i} 
    e^{\frac{K^{\ }_{A}}{2} J 
       \Sigma^{\ }_{\P^{\ }_{i}} \Theta^{(l)}_{i}}_{\ } 
\nonumber \\ 
&& 
\qquad
= 
\sum^{\ }_{J = \pm 1} 
  \left( 
    \prod^{\ }_{s} 
      \sum^{\ }_{\theta^{(l)}_{s} = \pm 1} 
        e^{\frac{K^{\ }_{A}}{2} J \theta^{(l)}_{s}}_{\ } 
  \right)
\times
\nonumber \\ 
&& 
\qquad\qquad\qquad
  \left( 
    \prod^{\ }_{i} 
      \sum^{\ }_{\Theta^{(l)}_{i} = \pm 1} 
        e^{\frac{K^{\ }_{A}}{2} J 
           \Sigma^{\ }_{\P^{\ }_{i}} \Theta^{(l)}_{i}}_{\ } 
  \right). 
\eea

Using the expression above, one can rewrite Eq.~\ref{eq: Tr(rho^n_A(T)) -- 1} 
as 
\begin{widetext}
\bea
\textrm{Tr} \left[ \rho^{n}_{\A} \right] 
&=& 
\left( \frac{d^{\ }_{\B}}{\vert G \vert} \right)^{n-1}_{\ } 
\frac{1}
     {2^{2n}_{\ } 
      \left[\cosh\left( \frac{K^{\ }_{A}}{2} N \right)\right]^{n}_{\ }
     } 
\: 
\prod^{n}_{l=1} 
\sum^{\ }_{J^{\ }_{l} = \pm 1} 
  \left( 
    \prod^{\ }_{s} 
      \sum^{\ }_{\theta^{(l)}_{s} = \pm 1} 
        e^{\frac{K^{\ }_{A}}{2} J^{\ }_{l} \theta^{(l)}_{s}}_{\ } 
  \right)
  \left( 
    \prod^{\ }_{i} 
      \sum^{\ }_{\Theta^{(l)}_{i} = \pm 1} 
        e^{\frac{K^{\ }_{A}}{2} J^{\ }_{l} 
           \Sigma^{\ }_{\P^{\ }_{i}} \Theta^{(l)}_{i}}_{\ } 
  \right) 
\times
\nonumber \\ 
&& 
\left\{ 
\prod^{\ }_{i,s} 
  \delta(\prod^{n}_{l=1} \Theta^{(l)}_{i} - 1) 
  \delta(\prod^{n}_{l=1} \theta^{(l)}_{s} - 1) 
+ 
\prod^{\ }_{i,s} 
  \delta(\prod^{n}_{l=1} \Theta^{(l)}_{i} + 1) 
  \delta(\prod^{n}_{l=1} \theta^{(l)}_{s} + 1) 
\right\} 
\nonumber 
\eea
%
%
and, upon expanding the product 
$
\prod^{n}_{l=1} 
  \left( 
    \sum^{\ }_{J^{\ }_{l} = \pm 1} 
      C(J^{\ }_{l}) 
  \right)
= 
\sum^{\ }_{\{ J^{\ }_{l} \}^{n}_{l=1}} 
  \prod^{n}_{l=1} 
    C(J^{\ }_{l})
$, 
we obtain 
\bea
\textrm{Tr} \left[ \rho^{n}_{\A} \right] 
&=& 
\left( \frac{d^{\ }_{\B}}{\vert G \vert} \right)^{n-1}_{\ } 
\frac{1}
     {2^{2n}_{\ } 
      \left[\cosh\left( \frac{K^{\ }_{A}}{2} N \right)\right]^{n}_{\ }
     } 
\: 
\sum^{\ }_{\{ J^{\ }_{l} \}^{n}_{l=1}} 
\left\{ 
  \left( 
    \prod^{\ }_{s} 
      \mathop{\sum^{\ }_{\{ \theta^{(l)}_{s} \}^{n}_{l=1}}}
             ^{\ }_{\prod \theta^{(l)}_{s} = +1} 
        e^{\frac{K^{\ }_{A}}{2} \sum^{\ }_{l} J^{\ }_{l} \theta^{(l)}_{s}}_{\ } 
  \right)
  \left( 
    \prod^{\ }_{i} 
      \mathop{\sum^{\ }_{\{ \Theta^{(l)}_{i} \}^{n}_{l=1}}}
             ^{\ }_{\prod \Theta^{(l)}_{i} = +1} 
        e^{\frac{K^{\ }_{A}}{2} \sum^{\ }_{l} J^{\ }_{l} 
           \Sigma^{\ }_{\P^{\ }_{i}} \Theta^{(l)}_{i}}_{\ } 
  \right) 
+ 
\right. 
\nonumber \\ 
&& 
\left. 
  \left( 
    \prod^{\ }_{s} 
      \mathop{\sum^{\ }_{\{ \theta^{(l)}_{s} \}^{n}_{l=1}}}
             ^{\ }_{\prod \theta^{(l)}_{s} = -1} 
        e^{\frac{K^{\ }_{A}}{2} \sum^{\ }_{l} J^{\ }_{l} \theta^{(l)}_{s}}_{\ } 
  \right)
  \left( 
    \prod^{\ }_{i} 
      \mathop{\sum^{\ }_{\{ \Theta^{(l)}_{i} \}^{n}_{l=1}}}
             ^{\ }_{\prod \Theta^{(l)}_{i} = +1} 
        e^{\frac{K^{\ }_{A}}{2} \sum^{\ }_{l} J^{\ }_{l} 
           \Sigma^{\ }_{\P^{\ }_{i}} \Theta^{(l)}_{i}}_{\ } 
  \right) 
\right\} 
\label{eq: Tr(rho^n_A(T)) -- 2}
\eea
\end{widetext} 
Notice that the indices $s$ and $i$ to the variables 
$\theta^{(l)}_{s}$ and $\Theta^{(l)}_{i}$, respectively, are mute since 
the sums over the possible values of $\theta^{(l)}_{s}$ and 
$\Theta^{(l)}_{i}$ are performed first. 
Therefore one can simplify the notation above by replacing 
$\theta^{(l)}_{s} \to \theta^{(l)}_{\ }$ and 
$\Theta^{(l)}_{i} \to \Theta^{(l)}_{\ }$ for all $s$ and $i$. 
Moreover, one can recognize that an Ising spin chain 
$\{ \theta^{(l)}_{\ } \}$ with $\prod \theta^{(l)}_{\ } = \pm 1$ is dual 
to an Ising spin chain $\{ \tau^{\ }_{l} \}$ with periodic / antiperiodic 
boundary conditions (using the $2$-to-$1$ mapping 
$\theta^{(l)}_{\ } = \tau^{\ }_{l} \tau^{\ }_{l+1}$). 
Thus, 
\bea
\mathop{\sum^{\ }_{\{ \theta^{(l)}_{\ } \}^{n}_{l=1}}}
       ^{\ }_{\prod \theta^{(l)}_{\ } = \pm 1} 
  e^{\frac{K^{\ }_{A}}{2} \sum^{n}_{l=1} J^{\ }_{l} \theta^{(l)}_{\ }}_{\ } 
&=& 
\frac{1}{2} 
\sum^{\textrm{p.} / \textrm{a.}}_{\{ \tau^{\ }_{l} \}^{n}_{l=1}}
  e^{\frac{K^{\ }_{A}}{2} \sum^{n}_{l=1} J^{\ }_{l} 
     \tau^{\ }_{l}\tau^{\ }_{l+1}}_{\ } 
\nonumber \\ 
&\equiv& 
\frac{1}{2} 
Z^{(p/a)}_{n}(K^{\ }_{A},\{J^{\ }_{l}\}), 
\eea
where $Z^{(p/a)}_{n}(K^{\ }_{A},\{J^{\ }_{l}\})$ is the partition function 
of a chain of $n$ Ising spins with periodic / antiperiodic boundary 
conditions, in presence of a nearest-neighbor interaction with 
position-dependent reduced coupling constant $K^{\ }_{A} J^{\ }_{l} / 2$. 
Similarly, 
\bea
\mathop{\sum^{\ }_{\{ \Theta^{(l)}_{\ } \}^{n}_{l=1}}}
       ^{\ }_{\prod \Theta^{(l)}_{\ } = \pm 1} 
  e^{\frac{K^{\ }_{A}}{2} \sum^{n}_{l=1} 
     \Sigma^{\ }_{\P^{\ }_{i}} J^{\ }_{l} \Theta^{(l)}_{\ }}_{\ } 
&\equiv& 
\frac{1}{2} 
Z^{(p/a)}_{n}(K^{\ }_{A} \Sigma^{\ }_{\P^{\ }_{i}},\{J^{\ }_{l}\}), 
\nonumber \\  
\eea
and Eq.~(\ref{eq: Tr(rho^n_A(T)) -- 2}) can be rewritten as 
\begin{widetext}
\bea
\textrm{Tr} \left[ \rho^{n}_{\A} \right] 
&=& 
\frac{1}{2^{\Sigma^{\ }_{\A}+m^{\ }_{\B}}_{\ }} 
\left( 
  \frac{d^{\ }_{\B}}{\vert G \vert} 
\right)^{n-1}_{\ } 
\frac{1}
     {
      2^{2n}_{\ } 
      \left[\cosh\left( \frac{K^{\ }_{A}}{2} N \right)\right]^{n}_{\ }
     } 
\: 
\sum^{\ }_{\{ J^{\ }_{l} \}^{n}_{l=1}} 
\times 
\nonumber \\ 
&& 
\left\{ 
  \left( 
    Z^{(p)}_{n}(K^{\ }_{A},\{ J^{\ }_{l} \})
  \right)^{\Sigma^{\ }_{\A}}_{\ } 
  \prod^{\ }_{i} 
  \left(\vphantom{\prod^{1}_{0}} 
      Z^{(p)}_{n}(K^{\ }_{A} \Sigma^{\ }_{\P^{\ }_{i}},\{J^{\ }_{l}\})
  \right) 
+ 
  \left( 
    Z^{(a)}_{n}(K^{\ }_{A},\{ J^{\ }_{l} \})
  \right)^{\Sigma^{\ }_{\A}}_{\ } 
  \prod^{\ }_{i} 
  \left(\vphantom{\prod^{1}_{0}} 
      Z^{(a)}_{n}(K^{\ }_{A} \Sigma^{\ }_{\P^{\ }_{i}},\{J^{\ }_{l}\})
  \right) 
\right\}. 
\nonumber \\ 
\label{eq: Tr(rho^n_A(T)) -- 3}
\eea
\end{widetext} 

The partition functions can be evaluated, for either boundary conditions, 
using a transfer matrix approach. 
For periodic boundary conditions, we obtain 
\beq
Z^{(p)}_{n}(K,\{J^{\ }_{l}\}) 
= 
\textrm{Tr} 
  \left[ 
    \prod^{n}_{l=1} T^{\ }_{l} 
  \right], 
\;\;\;\;\;\;
T^{\ }_{l} 
= 
\left( 
\begin{array}{cc} 
  e^{\frac{K}{2} J^{\ }_{l}}_{\ } & e^{-\frac{K}{2} J^{\ }_{l}}_{\ } 
  \\ 
  e^{-\frac{K}{2} J^{\ }_{l}}_{\ } & e^{\frac{K}{2} J^{\ }_{l}}_{\ } 
\end{array}
\right). 
\nonumber
\eeq
Since all matrices $T^{\ }_{l}$ are diagonalized by the same unitary matrix 
\beq
U 
= 
\frac{1}{\sqrt{2}} 
\left( 
\begin{array}{cc} 
  1 & 1 \\ 
  -1 & 1 
\end{array}
\right), 
\eeq
and using the cyclic properties of the trace, we get 
\bea
Z^{(p)}_{n}(K,\{J^{\ }_{l}\}) 
&=& 
\textrm{Tr} 
  \left[ 
    \prod^{n}_{l=1} (U T^{\ }_{l} U^{\dagger}_{\ })
  \right] 
\nonumber \\ 
&& 
\!\!\!\!\!\!\!\!\!\!\!\!\!\!\!\!\!\!\!\!\!\!\!\!\!\!\!
= 
\textrm{Tr} 
  \left[ 
    \prod^{n}_{l=1} 
      \left( 
        \begin{array}{cc} 
          e^{\frac{K}{2} J^{\ }_{l}}_{\ } + e^{-\frac{K}{2} J^{\ }_{l}}_{\ } 
	  & 
	  0 
	  \\ 
          0 
	  & 
	  e^{\frac{K}{2} J^{\ }_{l}}_{\ } - e^{-\frac{K}{2} J^{\ }_{l}}_{\ } 
        \end{array}
      \right) 
  \right] 
\nonumber \\ 
&& 
\!\!\!\!\!\!\!\!\!\!\!\!\!\!\!\!\!\!\!\!\!\!\!\!\!\!\!
= 
\textrm{Tr} 
  \left[ 
    \prod^{n}_{l=1} 
      \left( 
        \begin{array}{cc} 
          2 \cosh(\frac{K}{2}) 
	  & 
	  0 
	  \\ 
          0 
	  & 
          2 J^{\ }_{l} \sinh(\frac{K}{2})
        \end{array}
      \right) 
  \right] 
\nonumber \\ 
&& 
\!\!\!\!\!\!\!\!\!\!\!\!\!\!\!\!\!\!\!\!\!\!\!\!\!\!\!
= 
\left[ 2 \cosh\left(\frac{K}{2}\right) \right]^{n}_{\ } 
+ 
\left( \prod^{n}_{l=1} J^{\ }_{l} \right) 
\left[ 2 \sinh\left(\frac{K}{2}\right) \right]^{n}_{\ }. 
\nonumber \\ 
\eea
For the antiperiodic case, 
\beq
Z^{(a)}_{n}(K,\{J^{\ }_{l}\}) 
= 
\textrm{Tr} 
  \left[ 
    \sigma^{\ }_{1} 
    \prod^{n}_{l=1} T^{\ }_{l} 
  \right] 
\;\;\;\;\;\;\;\;
\sigma^{\ }_{1} 
= 
\left( 
\begin{array}{cc} 
  0 & 1 
  \\ 
  1 & 0 
\end{array}
\right), 
\nonumber
\eeq
and we get 
\bea
Z^{(a)}_{n}(K,\{J^{\ }_{l}\}) 
&=& 
\textrm{Tr} 
  \left[ 
    (U \sigma^{\ }_{1} U^{\dagger}_{\ })
    \prod^{n}_{l=1}(U T^{\ }_{l} U^{\dagger}_{\ })
  \right] 
\nonumber \\ 
&& 
\!\!\!\!\!\!\!\!\!\!\!\!\!\!\!\!\!\!\!\!\!\!\!\!\!\!\!
= 
\left[ 2 \cosh\left(\frac{K}{2}\right) \right]^{n}_{\ } 
- 
\left( \prod^{n}_{l=1} J^{\ }_{l} \right) 
\left[ 2 \sinh\left(\frac{K}{2}\right) \right]^{n}_{\ }. 
\nonumber \\ 
\eea

We can substitute into Eq.~(\ref{eq: Tr(rho^n_A(T)) -- 3}) and we obtain 
\begin{widetext}
\bea
\textrm{Tr} \left[ \rho^{n}_{\A} \right] 
&=& 
\left( 
  \frac{d^{\ }_{\B}\, 2^{\Sigma^{\ }_{\A}+m^{\ }_{\B}}_{\ }}{\vert G \vert} 
\right)^{n-1}_{\ } 
\frac{1}
     {
      2^{2n}_{\ } 
      \left[\cosh\left( \frac{K^{\ }_{A}}{2} N \right)\right]^{n}_{\ }
     } 
\: 
\sum^{\ }_{\{ J^{\ }_{l} \}^{n}_{l=1}} 
\times 
\nonumber \\ 
&& 
\!\!\!\!\!\!\!\!\!\!\!\!\!\!\!\!\!\!\!\!\!\!\!\!
\left\{ 
  \left( 
    \left[ \cosh\left(\frac{K^{\ }_{A}}{2}\right) \right]^{n}_{\ } 
    + 
    \left( \prod^{n}_{l=1} J^{\ }_{l} \right) 
    \left[ \sinh\left(\frac{K^{\ }_{A}}{2}\right) \right]^{n}_{\ } 
  \right)^{\Sigma^{\ }_{\A}}_{\ } 
  \prod^{\ }_{i} 
  \left(\vphantom{\prod^{1}_{0}} 
    \left[ \cosh\left(\frac{K^{\ }_{A}}{2} \Sigma^{\ }_{\P^{\ }_{i}}\right) \right]^{n}_{\ } 
    + 
    \left( \prod^{n}_{l=1} J^{\ }_{l} \right) 
    \left[ \sinh\left(\frac{K^{\ }_{A}}{2} \Sigma^{\ }_{\P^{\ }_{i}}\right) \right]^{n}_{\ } 
  \right) 
\right. 
\nonumber \\ 
&& 
\!\!\!\!\!\!\!\!\!\!\!\!\!\!\!\!\!\!\!\!\!\!\!\!
\left. 
  + 
  \left( 
    \left[ \cosh\left(\frac{K^{\ }_{A}}{2}\right) \right]^{n}_{\ } 
    - 
    \left( \prod^{n}_{l=1} J^{\ }_{l} \right) 
    \left[ \sinh\left(\frac{K^{\ }_{A}}{2}\right) \right]^{n}_{\ } 
  \right)^{\Sigma^{\ }_{\A}}_{\ } 
  \prod^{\ }_{i} 
  \left(\vphantom{\prod^{1}_{0}} 
    \left[ \cosh\left(\frac{K^{\ }_{A}}{2} \Sigma^{\ }_{\P^{\ }_{i}}\right) \right]^{n}_{\ } 
    - 
    \left( \prod^{n}_{l=1} J^{\ }_{l} \right) 
    \left[ \sinh\left(\frac{K^{\ }_{A}}{2} \Sigma^{\ }_{\P^{\ }_{i}}\right) \right]^{n}_{\ } 
  \right) 
\right\}. 
\nonumber 
\eea
Notice that the factors $\prod^{\ }_{l} J^{\ }_{l}$ can be dropped since the 
two terms between curly brackets in the equation above get simply exchanged 
by the two different values of $\prod^{\ }_{l} J^{\ }_{l} = \pm 1$. 
The summation over $\{ J^{\ }_{l} \}^{n}_{l=1}$ gives therefore just a 
multiplicative factor $2^{n}_{\ }$. 
Recalling the definition 
$d^{\ }_{\A} = 2^{\Sigma^{\ }_{\A}+m^{\ }_{\B}-1}_{\ }$, we finally get to 
\bea
\textrm{Tr} \left[ \rho^{n}_{\A} \right] 
&=& 
\frac{1}{2 \cosh\left( \frac{K^{\ }_{A}}{2} N \right)} 
\left( 
  \frac{1}{\cosh\left( \frac{K^{\ }_{A}}{2} N \right)} 
  \frac{d^{\ }_{\A} d^{\ }_{\B}}{\vert G \vert} 
\right)^{n-1}_{\ } 
\times 
\nonumber \\ 
&& 
\left\{ 
  \left( 
    \left[ \cosh\left(\frac{K^{\ }_{A}}{2}\right) \right]^{n}_{\ } 
    + 
    \left[ \sinh\left(\frac{K^{\ }_{A}}{2}\right) \right]^{n}_{\ } 
  \right)^{\Sigma^{\ }_{\A}}_{\ } 
  \prod^{\ }_{i} 
  \left(\vphantom{\prod^{1}_{0}} 
    \left[ \cosh\left(\frac{K^{\ }_{A}}{2} \Sigma^{\ }_{\P^{\ }_{i}}\right) \right]^{n}_{\ } 
    + 
    \left[ \sinh\left(\frac{K^{\ }_{A}}{2} \Sigma^{\ }_{\P^{\ }_{i}}\right) \right]^{n}_{\ } 
  \right) 
+ 
\right. 
\nonumber \\ 
&& 
\left. 
  \left( 
    \left[ \cosh\left(\frac{K^{\ }_{A}}{2}\right) \right]^{n}_{\ } 
    - 
    \left[ \sinh\left(\frac{K^{\ }_{A}}{2}\right) \right]^{n}_{\ } 
  \right)^{\Sigma^{\ }_{\A}}_{\ } 
  \prod^{\ }_{i} 
  \left(\vphantom{\prod^{1}_{0}} 
    \left[ \cosh\left(\frac{K^{\ }_{A}}{2} \Sigma^{\ }_{\P^{\ }_{i}}\right) \right]^{n}_{\ } 
    - 
    \left[ \sinh\left(\frac{K^{\ }_{A}}{2} \Sigma^{\ }_{\P^{\ }_{i}}\right) \right]^{n}_{\ } 
  \right) 
\right\}. 
\label{eq: Tr(rho^n_A(T)) -- 4}
\eea
\end{widetext} 

We can now use the identity 
\beq
\textrm{Tr} \left[ \rho^{\ }_{\A} \ln \rho^{\ }_{\A} \right]
= 
\lim^{\ }_{n \to 1} 
  \partial^{\ }_{n} \textrm{Tr} \left[ \rho^{n}_{\A} \right] 
\eeq
to compute the von Neumann entropy 
$
S^{\A}_{VN}(T) 
= 
- 
\textrm{Tr} \left[ \rho^{\ }_{\A} \ln \rho^{\ }_{\A} \right]
$. 
It is convenient to introduce the following simplified notation: 
\begin{subequations}
\beq
x = \cosh\left( \frac{K^{\ }_{A}}{2} \right)
\qquad 
y = \sinh\left( \frac{K^{\ }_{A}}{2} \right)
\eeq
\beq
\tilde{x}^{\ }_{i} 
= 
\cosh\left( \frac{K^{\ }_{A}}{2} \Sigma^{\ }_{\P^{\ }_{i}} \right)
\qquad 
\tilde{y}^{\ }_{i} 
= 
\sinh\left( \frac{K^{\ }_{A}}{2} \Sigma^{\ }_{\P^{\ }_{i}} \right), 
\eeq
\end{subequations}
which allows us to rewrite the terms in the last two lines of 
Eq.~(\ref{eq: Tr(rho^n_A(T)) -- 4}) as 
\bea
F^{(n)}_{+} 
&=& 
\left( x^{n}_{\ } + y^{n}_{\ } \right)^{\Sigma^{\ }_{\A}}_{\ } 
\prod^{\ }_{i} 
\left( \tilde{x}^{n}_{i} + \tilde{y}^{n}_{i} \right) 
\\ 
F^{(n)}_{-} 
&=& 
\left( x^{n}_{\ } - y^{n}_{\ } \right)^{\Sigma^{\ }_{\A}}_{\ } 
\prod^{\ }_{i} 
\left( \tilde{x}^{n}_{i} - \tilde{y}^{n}_{i} \right). 
\eea

One can verify that 
\bea
F^{(1)}_{\pm} 
&=& 
e^{
   \pm\frac{K^{\ }_{A}}{2} 
     \left( 
       \Sigma^{\ }_{\A} 
       +
       \sum^{\ }_{i} \Sigma^{\ }_{\P^{\ }_{i}}
     \right)
  }_{\ } 
\nonumber \\ 
&=& 
e^{\pm\frac{K^{\ }_{A}}{2} N}_{\ } 
\eea
and
\bea 
\left. 
  \partial^{\ }_{n} F^{(n)}_{\pm} 
\vphantom{\int}\right\vert^{\ }_{n=1} 
&=& 
\Sigma^{\ }_{\A} 
e^{\pm\frac{K^{\ }_{A}}{2} \left( N - 1 \right)}_{\ } 
\left( x \ln x \pm y \ln y \right) 
\nonumber \\ 
&& 
+ 
\sum^{\ }_{i} 
  e^{
     \pm\frac{K^{\ }_{A}}{2} \left( N - \Sigma^{\ }_{\P^{\ }_{i}} \right) 
    }_{\ } 
    \left( 
      \tilde{x}^{\ }_{i} \ln \tilde{x}^{\ }_{i} 
      \pm 
      \tilde{y}^{\ }_{i} \ln \tilde{y}^{\ }_{i} 
    \right). 
\nonumber \\ 
\eea

\begin{widetext}
Using this notation 
\bea
\textrm{Tr} \left[ \rho^{n}_{\A} \right] 
&=& 
\frac{1}
     {2 \cosh\left( \frac{K^{\ }_{A}}{2} N \right)} 
\left( 
  \frac{1}{\cosh\left( \frac{K^{\ }_{A}}{2} N \right)} 
  \frac{d^{\ }_{\A} d^{\ }_{\B}}{\vert G \vert} 
\right)^{n-1}_{\ } 
\left\{ F^{(n)}_{+} + F^{(n)}_{-} \right\}. 
\nonumber 
\eea
and 
\bea
S^{\A}_{\textrm{VN}}(T) 
&=& 
- 
\frac{1}
     {2 \cosh\left( \frac{K^{\ }_{A}}{2} N \right)} 
\left\{
  \ln\left[ 
       \frac{1}{\cosh\left( \frac{K^{\ }_{A}}{2} N \right)} 
       \frac{d^{\ }_{\A} d^{\ }_{\B}}{\vert G \vert} 
     \right]
  2 \cosh\left( \frac{K^{\ }_{A}}{2} N \right) 
\right.
\nonumber \\ 
&& 
\qquad\qquad
\left. 
+ 
\Sigma^{\ }_{\A} 
e^{\frac{K^{\ }_{A}}{2} \left( N - 1 \right)}_{\ } 
\left( x \ln x + y \ln y \right) 
+ 
\sum^{\ }_{i} 
  e^{
     \frac{K^{\ }_{A}}{2} \left( N - \Sigma^{\ }_{\P^{\ }_{i}} \right) 
    }_{\ } 
    \left( 
      \tilde{x}^{\ }_{i} \ln \tilde{x}^{\ }_{i} 
      + 
      \tilde{y}^{\ }_{i} \ln \tilde{y}^{\ }_{i} 
    \right) 
\right.
\nonumber \\ 
&& 
\qquad\qquad
\left. 
+ 
\Sigma^{\ }_{\A} 
e^{-\frac{K^{\ }_{A}}{2} \left( N - 1 \right)}_{\ } 
\left( x \ln x - y \ln y \right) 
+ 
\sum^{\ }_{i} 
  e^{
     -\frac{K^{\ }_{A}}{2} \left( N - \Sigma^{\ }_{\P^{\ }_{i}} \right) 
    }_{\ } 
    \left( 
      \tilde{x}^{\ }_{i} \ln \tilde{x}^{\ }_{i} 
      - 
      \tilde{y}^{\ }_{i} \ln \tilde{y}^{\ }_{i} 
    \right) 
\right\}
\nonumber \\ 
&=& 
- 
\ln \frac{d^{\ }_{\A} d^{\ }_{\B}}{\vert G \vert} 
+ 
\ln \cosh\left( \frac{K^{\ }_{A}}{2} N \right) 
\nonumber \\ 
&& 
- 
\frac{1}
     {2 \cosh\left( \frac{K^{\ }_{A}}{2} N \right)} 
\left\{
\Sigma^{\ }_{\A} 
e^{\frac{K^{\ }_{A}}{2} \left( N - 1 \right)}_{\ } 
\left( x \ln x + y \ln y \right) 
+ 
\sum^{\ }_{i} 
  e^{
     \frac{K^{\ }_{A}}{2} \left( N - \Sigma^{\ }_{\P^{\ }_{i}} \right) 
    }_{\ } 
    \left( 
      \tilde{x}^{\ }_{i} \ln \tilde{x}^{\ }_{i} 
      + 
      \tilde{y}^{\ }_{i} \ln \tilde{y}^{\ }_{i} 
    \right) 
\right.
\nonumber \\ 
&& 
\left. 
+ 
\Sigma^{\ }_{\A} 
e^{-\frac{K^{\ }_{A}}{2} \left( N - 1 \right)}_{\ } 
\left( x \ln x - y \ln y \right) 
+ 
\sum^{\ }_{i} 
  e^{
     -\frac{K^{\ }_{A}}{2} \left( N - \Sigma^{\ }_{\P^{\ }_{i}} \right) 
    }_{\ } 
    \left( 
      \tilde{x}^{\ }_{i} \ln \tilde{x}^{\ }_{i} 
      - 
      \tilde{y}^{\ }_{i} \ln \tilde{y}^{\ }_{i} 
    \right) 
\right\}
\nonumber \\ 
&=& 
- 
\ln \frac{d^{\ }_{\A} d^{\ }_{\B}}{\vert G \vert} 
+ 
\ln \cosh\left( \frac{K^{\ }_{A}}{2} N \right) 
- 
\Sigma^{\ }_{\A} \left( x \ln x \right) 
  \frac{\cosh\left( \frac{K^{\ }_{A}}{2} (N - 1) \right)}
       {\cosh\left( \frac{K^{\ }_{A}}{2} N \right)}
- 
\Sigma^{\ }_{\A} \left( y \ln y \right) 
  \frac{\sinh\left( \frac{K^{\ }_{A}}{2} (N - 1) \right)}
       {\cosh\left( \frac{K^{\ }_{A}}{2} N \right)}
\nonumber \\ 
&& 
- 
\sum^{\ }_{i} 
\left( \tilde{x}^{\ }_{i} \ln \tilde{x}^{\ }_{i} \right) 
  \frac{\cosh\left( \frac{K^{\ }_{A}}{2} (N - \Sigma^{\ }_{\P^{\ }_{i}}) \right)}
       {\cosh\left( \frac{K^{\ }_{A}}{2} N \right)} 
- 
\sum^{\ }_{i} 
\left( \tilde{y}^{\ }_{i} \ln \tilde{y}^{\ }_{i} \right) 
  \frac{\sinh\left( \frac{K^{\ }_{A}}{2} (N - \Sigma^{\ }_{\P^{\ }_{i}}) \right)}
       {\cosh\left( \frac{K^{\ }_{A}}{2} N \right)} 
\label{eq: S^A_VN(T)}
\eea
\end{widetext} 

For a finite size system, the limit of $K^{\ }_{A} \to 0^{+}_{\ }$ yields 
$x,\tilde{x}^{\ }_{i} \to 1$, $y,\tilde{y}^{\ }_{i} \to 0$ and 
therefore 
\beq
S^{\A}_{\textrm{VN}}(T \to 0) 
\rightarrow 
- \ln \frac{d^{\ }_{\A} d^{\ }_{\B}}{\vert G \vert}, 
\label{eq: S_VN(N,T -> 0)}
\eeq
consistently with the known zero-temperature result.~\cite{Hamma2005} 

In the limit of $K^{\ }_{A} \to \infty$ instead, 
$x,y \sim e^{K^{\ }_{A}/2}_{\ } / 2$, 
$
\tilde{x}^{\ }_{i}, \tilde{y}^{\ }_{i} 
\sim 
e^{K^{\ }_{A} \Sigma^{\ }_{\P^{\ }_{i}}/2}_{\ } / 2
$, 
and therefore 
\bea
&& 
S^{\A}_{\textrm{VN}}(T \to \infty) 
\rightarrow 
\nonumber \\
&& 
\qquad
\rightarrow 
- 
\ln \frac{d^{\ }_{\A} d^{\ }_{\B}}{\vert G \vert} 
+ 
\frac{K^{\ }_{A}}{2} N 
- 
\ln 2 
- 
\Sigma^{\ }_{\A} 
  \left( \frac{K^{\ }_{A}}{2} - \ln 2 \right) 
\nonumber \\
&& 
\qquad\qquad
- 
\sum^{\ }_{i} 
  \left( 
    \frac{K^{\ }_{A}}{2} \Sigma^{\ }_{\P^{\ }_{i}} 
    - \ln 2 
  \right) 
\nonumber \\ 
&& 
\qquad
= 
- 
\ln \frac{d^{\ }_{\A} d^{\ }_{\B}}{\vert G \vert} 
- 
\ln 2 
+ 
\Sigma^{\ }_{\A} \ln 2 
+ 
m^{\ }_{\B} \ln 2 
\nonumber \\ 
&& 
\qquad
= 
- \ln \frac{d^{\ }_{\B}}{\vert G \vert}, 
\eea
where $d^{\ }_{\A} = 2^{\Sigma^{\ }_{\A} + m^{\ }_{\B} - 1}_{\ }$. 
This result is consistent with Ref.~\onlinecite{Castelnovo2006}, where 
the classical system that obtains in the $T \to \infty$ limit had been 
previously discussed. 

Let us then consider the thermodynamic limit $L \to \infty$. 
The presence of terms that depend on the product between 
$K^{\ }_{A}$ and extensive quantities such 
as $N$, $\Sigma^{\ }_{\A}$, and $\Sigma^{\ }_{\P^{\ }_{i}}$ requires 
careful consideration. As we will see, they will indeed lead to a singular 
behavior of the von Neumann entropy, as well as the topological entropy 
discussed in Sec.~\ref{sec: topo entropy}. 
For any finite $K^{\ }_{A} \in (0,\infty)$, the limit $L \to \infty$ 
($N = L^{2}_{\ } \to \infty$) yields 
\bea
&& 
S^{\A}_{\textrm{VN}}(T) 
\stackrel{L \to \infty}{\longrightarrow} 
\nonumber \\
&& 
\qquad
\longrightarrow 
- 
\ln \frac{d^{\ }_{\A} d^{\ }_{\B}}{\vert G \vert} 
+ 
\frac{K^{\ }_{A}}{2} N 
- 
\ln 2 
\nonumber \\
&& 
\qquad\qquad
- 
\Sigma^{\ }_{\A} 
  \left( 
    x \ln x 
    + 
    y \ln y 
  \right) 
    e^{-\frac{K^{\ }_{A}}{2}}_{\ } 
\nonumber \\
&& 
\qquad\qquad
- 
\sum^{\ }_{i} 
  \left( 
    \tilde{x}^{\ }_{i} \ln \tilde{x}^{\ }_{i} 
    + 
    \tilde{y}^{\ }_{i} \ln \tilde{y}^{\ }_{i} 
  \right) 
    e^{-\frac{K^{\ }_{A}}{2} \Sigma^{\ }_{\P^{\ }_{i}}}_{\ }. 
\label{eq: S_VN(T) for large L no Sigma} 
\eea
We can further simplify this expression in the limit 
$\Sigma^{\ }_{\P^{\ }_{i}}, \Sigma^{\ }_{\A} \gg 1$, 
i.e., in the limit of large partitions;~\cite{footnote: large partitions} 
this is the case of interest, for example, in the definition of the 
topological entropy discussed in the Section~\ref{sec: topo entropy}. 
If we expand 
$
\tilde{x}^{\ }_{i}, \tilde{y}^{\ }_{i} 
\sim 
e^{K^{\ }_{A} \Sigma^{\ }_{\P^{\ }_{i}}/2}_{\ } / 2
$, 
we obtain 
\bea
&& 
S^{\A}_{\textrm{VN}}(T) 
\stackrel{L \to \infty}{\longrightarrow} 
\nonumber \\
&& 
\qquad
\longrightarrow 
- 
\ln \frac{d^{\ }_{\A} d^{\ }_{\B}}{\vert G \vert} 
+ 
\frac{K^{\ }_{A}}{2} N 
- 
\ln 2 
\nonumber \\
&& 
\qquad\qquad
- 
\Sigma^{\ }_{\A} e^{-\frac{K^{\ }_{A}}{2}}_{\ }
  \left( 
    x \ln x 
    + 
    y \ln y 
  \right) 
\nonumber \\
&& 
\qquad\qquad
- 
\sum^{\ }_{i} 
  \left( 
    \frac{K^{\ }_{A}}{2} \Sigma^{\ }_{\P^{\ }_{i}} 
    - \ln 2 
  \right) 
\nonumber \\ 
&& 
\qquad
= 
- 
\ln \frac{d^{\ }_{\A} d^{\ }_{\B}}{\vert G \vert} 
+ 
\left( m^{\ }_{\B} - 1 \right) \ln 2 
\nonumber \\
&& 
\qquad \;\;\;\; 
- 
\Sigma^{\ }_{\A} 
\left[ 
e^{-\frac{K^{\ }_{A}}{2}}_{\ }
  \left( 
    x \ln x 
    + 
    y \ln y 
  \right) 
- 
\frac{K^{\ }_{A}}{2} 
\right] 
\nonumber \\ 
&& 
\qquad
= 
\left(\vphantom{\sum} 
  \Sigma^{\ }_{\A\B} - m^{\ }_{\A} 
\right) 
  \ln 2 
\nonumber \\
&& 
\qquad \;\;\;\; 
- 
\Sigma^{\ }_{\A} 
\left[ 
e^{-\frac{K^{\ }_{A}}{2}}_{\ }
  \left( 
    x \ln x 
    + 
    y \ln y 
  \right) 
- 
\frac{K^{\ }_{A}}{2} 
\right], 
\nonumber \\ 
\label{eq: S_VN(T) for large L}
\eea
which is consistent with the limit 
$
S^{\A}_{\textrm{VN}}(T \to \infty) 
= 
- \ln (d^{\ }_{\B} / \vert G \vert)
$, 
but \emph{no longer consistent} with the finite size result 
$
S^{\A}_{\textrm{VN}}(T \to 0) 
= 
- \ln (d^{\ }_{\A} d^{\ }_{\B} / \vert G \vert)
$. 
In fact, 
\bea
\lim^{\ }_{T \to 0, L \to \infty} 
S^{\A}_{\textrm{VN}}(T) 
&=& 
- 
\ln \frac{d^{\ }_{\A} d^{\ }_{\B}}{\vert G \vert} 
+ 
\left( m^{\ }_{\B} - 1 \right) \ln 2, 
\nonumber \\ 
&=& 
\left(\vphantom{\sum} 
  \Sigma^{\ }_{\A\B} 
  - 
  m^{\ }_{\A} 
\right) 
  \ln 2 
\label{eq: S_VN T,L -> infty}
\eea
while 
\bea
\lim^{\ }_{L \to \infty, T \to 0} 
S^{\A}_{\textrm{VN}}(T) 
&=& 
- 
\ln \frac{d^{\ }_{\A} d^{\ }_{\B}}{\vert G \vert}. 
\nonumber \\ 
&=& 
\left(\vphantom{\sum} 
  \Sigma^{\ }_{\A\B} 
  - 
  m^{\ }_{\A} - m^{\ }_{\B} + 1 
\right) 
  \ln 2 
\nonumber \\ 
\label{eq: S_VN L,T -> infty}
\eea
Notice that already in Eq.~(\ref{eq: S_VN(T) for large L no Sigma}) the 
limit $T \to 0$ fails to give the known result, 
Eq.~(\ref{eq: S_VN(N,T -> 0)}). 
However, it is only in the assumption that all partitions scale with the 
system size that we arrive at Eq.~(\ref{eq: S_VN(T) for large L}). 
Notice also that the difference between the two orders of limits arises only 
if subsystem $\B$ has more than one connected component. Moreover, such 
difference is of order one, while the common term 
$-\ln (d^{\ }_{\A} d^{\ }_{\B}) / \vert G \vert$ 
scales linearly with the size of the boundary between $\A$ and $\B$ 
($\propto \Sigma^{\ }_{\A\B}$). 
Therefore, from a thermodynamic point of view the order of limits is 
immaterial to the von Neumann entanglement entropy. 
On the other hand, we will see in the following section how this difference 
plays a crucial role in the topological entropy of the system, and gives 
rise to a singular behavior at zero temperature in the thermodynamic 
limit. 
%
%

\subsection{\label{sec: mutual information}
The mutual information
           }
In the previous section, and in particular from Eq.~(\ref{eq: S^A_VN(T)}) 
we clearly see that the von Neumann entropy is symmetric upon exchange of 
$\A$ and $\B$, and it satisfies the so-called area law only at zero 
temperature (and for topologically ordered systems, symmetry is lost unless 
the $T \to 0$ limit is taken before the $L \to \infty$ limit). 
At any finite temperature, $S^{\A}_{\textrm{VN}}(T)$ acquires an extensive 
contribution scaling with the number of degrees of freedom in partition $\A$. 
In this sense, the von Neumann entropy ceases to be 
a good measure of the entropy contained in the boundary between the two 
subsystems, and therefore is no longer a good measure of entanglement at 
finite temperature. 

Alternatively, one can consider another quantity called the mutual 
information $I^{\ }_{\A\B}$, which we redefine for convenience multiplied 
by a factor $1/2$, 
\bea
I^{\ }_{\A\B}(T) 
&=& 
\frac{1}{2}
  \left[\vphantom{\sum} 
    S^{\A}_{\textrm{VN}}(T) 
    +
    S^{\B}_{\textrm{VN}}(T)
    -
    S^{\A \cup \B}_{\textrm{VN}}(T)
  \right], 
\nonumber \\
\eea
so that 
$I^{\ }_{\A\B}(0,N) = S^{\A}_{\textrm{VN}}(0,N) = S^{\B}_{\textrm{VN}}(0,N)$. 
Substituting the result from Eq.~(\ref{eq: S^A_VN(T)}), and recalling that 
$d^{\ }_{\A\cup\B} = \vert G \vert$, $d^{\ }_{\emptyset} = 1$ and 
$\Sigma^{\ }_{\A\cup\B} = N$, we obtain the 
behavior of $I^{\ }_{\A\B}(T)$ as a function of 
temperature and system size for a topologically ordered system, 
\begin{widetext}
\bea
I^{\ }_{\A\B}(T) 
&=& 
- 
\ln \frac{d^{\ }_{\A} d^{\ }_{\B}}{\vert G \vert} 
+ 
\frac{1}{2} 
\ln \cosh\left( \frac{K^{\ }_{A}}{2} N \right) 
+ 
\frac{\Sigma^{\ }_{\A\B}}{2} 
\left\{
\left( x \ln x \right) 
  \frac{\cosh\left( \frac{K^{\ }_{A}}{2} (N - 1) \right)}
       {\cosh\left( \frac{K^{\ }_{A}}{2} N \right)}
+ 
\left( y \ln y \right) 
  \frac{\sinh\left( \frac{K^{\ }_{A}}{2} (N - 1) \right)}
       {\cosh\left( \frac{K^{\ }_{A}}{2} N \right)}
\right\} 
\nonumber \\ 
&& 
- 
\frac{1}{2} 
\sum^{\ }_{i} 
\left( \tilde{x}^{\ }_{i} \ln \tilde{x}^{\ }_{i} \right) 
  \frac{\cosh\left( \frac{K^{\ }_{A}}{2} 
        (N - \Sigma^{\ }_{\B^{\ }_{i}} - \Sigma^{\ }_{\A\B^{\ }_{i}}) \right)}
       {\cosh\left( \frac{K^{\ }_{A}}{2} N \right)} 
- 
\frac{1}{2} 
\sum^{\ }_{i} 
\left( \tilde{y}^{\ }_{i} \ln \tilde{y}^{\ }_{i} \right) 
  \frac{\sinh\left( \frac{K^{\ }_{A}}{2} 
        (N - \Sigma^{\ }_{\B^{\ }_{i}} - \Sigma^{\ }_{\A\B^{\ }_{i}}) \right)}
       {\cosh\left( \frac{K^{\ }_{A}}{2} N \right)} 
\nonumber \\ 
&& 
- 
\frac{1}{2} 
\sum^{\ }_{i} 
\left( \tilde{x}^{\ }_{i} \ln \tilde{x}^{\ }_{i} \right) 
  \frac{\cosh\left( \frac{K^{\ }_{A}}{2} 
        (N - \Sigma^{\ }_{\A^{\ }_{i}} - \Sigma^{\ }_{\A\B^{\ }_{i}}) \right)}
       {\cosh\left( \frac{K^{\ }_{A}}{2} N \right)} 
- 
\frac{1}{2} 
\sum^{\ }_{i} 
\left( \tilde{y}^{\ }_{i} \ln \tilde{y}^{\ }_{i} \right) 
  \frac{\sinh\left( \frac{K^{\ }_{A}}{2} 
        (N - \Sigma^{\ }_{\A^{\ }_{i}} - \Sigma^{\ }_{\A\B^{\ }_{i}}) \right)}
       {\cosh\left( \frac{K^{\ }_{A}}{2} N \right)}. 
\label{eq: I_AB(T)}
\eea
\end{widetext}
The mutual information has the immediate advantage over the von Neumann 
entropy that it is symmetric upon exchange of subsystem $\A$ with $\B$. 
Moreover, while the general expression above seems to have an explicit 
dependence on the bulk degrees of freedom ($\Sigma^{\ }_{\A^{\ }_{i}}$ and 
$\Sigma^{\ }_{\B^{\ }_{i}}$), we show hereafter that in the 
thermodynamic limit as well as in the zero-temperature limit the mutual 
information depends only on the boundary degrees of freedom 
($\Sigma^{\ }_{\A\B}$) and, as such, 
is a good candidate for a measure of entanglement at finite as well as 
zero temperature. 

In the $T \to 0$ limit we recover the known result 
\bea
I^{\ }_{\A\B}(0) 
&=& 
- 
\ln\frac{d^{\ }_{\A} d^{\ }_{\B}}{\vert G \vert} 
\nonumber \\ 
&=& 
\left(\vphantom{\sum} 
  \Sigma^{\ }_{\A\B} - m^{\ }_{\A} - m^{\ }_{\B} + 1
\right) 
  \ln 2. 
\eea
In the $T \to \infty$ limit instead we obtain 
\bea
I^{\ }_{\A\B}(\infty) 
&=& 
- \frac{1}{2} 
\ln\frac{d^{\ }_{\A} d^{\ }_{\B}}{\vert G \vert} 
\nonumber \\ 
&=& 
\frac{1}{2} 
\left(\vphantom{\sum} 
  \Sigma^{\ }_{\A\B} - m^{\ }_{\A} - m^{\ }_{\B} + 1
\right) 
  \ln 2, 
\eea
which again shows only a dependence on boundary degrees of freedom, with 
the addition of topological terms of order one, and is symmetric under the 
exchange of $\A$ and $\B$. 

We can then compare these finite-size results with the behavior of 
$I^{\ }_{\A\B}(T)$ in the thermodynamic limit $L \to \infty$ for large 
partitions, 
\bea
I^{\ }_{\A\B}(T) 
&\simeq& 
- 
\ln\frac{d^{\ }_{\A} d^{\ }_{\B}}{\vert G \vert} 
\nonumber \\ 
&&
+ 
\frac{\Sigma^{\ }_{\A\B}}{2} 
\left\{ 
  \left(
    x \ln x 
    + 
    y \ln y 
  \right) 
    e^{-\frac{K^{\ }_{A}}{2}}_{\ } 
  - 
  \frac{K^{\ }_{A}}{2} 
\right\} 
\nonumber \\ 
&&
+ 
\frac{1}{2} 
\left(\vphantom{\sum} 
  m^{\ }_{\A} + m^{\ }_{\B} - 1 
\right) 
  \ln 2 
\nonumber \\ 
&=& 
\frac{\Sigma^{\ }_{\A\B}}{2} 
\left\{ 
  2 \ln 2 
  + 
  \left(
    x \ln x 
    + 
    y \ln y 
  \right) 
    e^{-\frac{K^{\ }_{A}}{2}}_{\ } 
  - 
  \frac{K^{\ }_{A}}{2} 
\right\} 
\nonumber \\ 
&&
- 
\frac{1}{2} 
\left(\vphantom{\sum} 
  m^{\ }_{\A} + m^{\ }_{\B} - 1 
\right) 
  \ln 2. 
\eea
Once again this is consistent with the finite size limit $T \to \infty$, but 
\emph{not} with the $T \to 0$ limit. In fact, 
\bea
\lim^{\ }_{L \to \infty, T \to 0} I^{\ }_{\A\B}(T) 
&=& 
-\ln\frac{d^{\ }_{\A} d^{\ }_{\B}}{\vert G \vert} 
\nonumber \\ 
&=& 
\left(\vphantom{\sum} 
  \Sigma^{\ }_{\A\B} - m^{\ }_{\A} - m^{\ }_{\B} + 1 
\right) 
  \ln 2 
\nonumber \\ 
\eea
while 
\bea
\lim^{\ }_{T \to 0, L \to \infty} I^{\ }_{\A\B}(T) 
&=& 
-\ln\frac{d^{\ }_{\A} d^{\ }_{\B}}{\vert G \vert} 
+ 
\frac{1}{2} 
\left(\vphantom{\sum} 
  m^{\ }_{\A} + m^{\ }_{\B} - 1 
\right) 
  \ln 2 
\nonumber \\ 
&=& 
\left[\vphantom{\sum} 
  \Sigma^{\ }_{\A\B} 
  - 
  \frac{1}{2} 
  \left(\vphantom{\sum} 
    m^{\ }_{\A} + m^{\ }_{\B} - 1 
  \right) 
\right] 
  \ln 2. 
\nonumber \\ 
\eea
The difference between the mutual information obtained for the two
order of limits gives
\bea
\Delta I ^{\ }_{\A\B}
&=& 
\lim^{\ }_{T \to 0, L \to \infty} I^{\ }_{\A\B}(T) 
- 
\lim^{\ }_{L \to \infty, T \to 0} I^{\ }_{\A\B}(T) 
\nonumber \\ 
&=& 
\frac{1}{2} 
\left(\vphantom{\sum} 
  m^{\ }_{\A} + m^{\ }_{\B} - 1
\right) 
  \ln 2. 
\eea
It is noteworthy that the boundary contribution drops out from $\Delta
I ^{\ }_{\A\B}$, but that this quantity still has an ${\cal O}(1)$
topological contribution that depends on the total number of
disconnected regions $m^{\ }_{\A} + m^{\ }_{\B} $ of partitions $\A$
and $\B$. Hence a topological contribution to the entanglement entropy
can be filtered out directly from a single bipartition using $\Delta I
^{\ }_{\A\B}$, in contrast to the constructions in
Refs.~\onlinecite{Levin2006,Kitaev2006} that require a linear
combination over multiple bipartitions.

%

\section{\label{sec: topo entropy}
The topological entropy
        }
\begin{figure}[ht]
\vspace{0.2 cm}
\includegraphics[width=0.98\columnwidth]{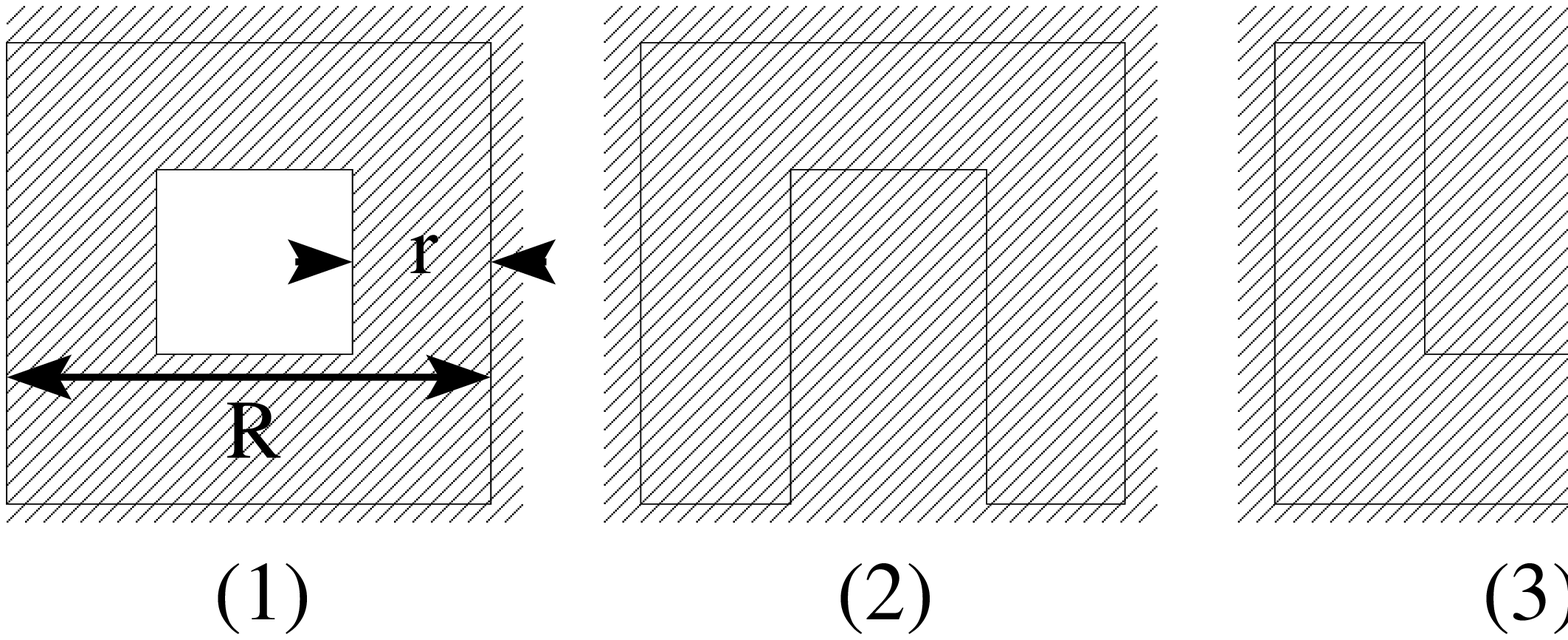}
\caption{
\label{fig: topological partitions}
Illustration of the four bipartitions used to compute the topological 
entropy in Ref.~\onlinecite{Levin2006}. 
}
\end{figure}

Let us now compute the topological entropy using the results for the 
von Neumann entropy in the previous section, and the definition given by 
Levin and Wen~\cite{Levin2006} 
\beq
S^{\ }_{\textrm{topo}} 
= 
\lim^{\ }_{r,R \to \infty} 
\left[ 
  - S^{1\A}_{\textrm{VN}} 
  + S^{2\A}_{\textrm{VN}} 
  + S^{3\A}_{\textrm{VN}} 
  - S^{4\A}_{\textrm{VN}} 
\right] 
\eeq
based on the bipartitions shown in 
Fig.~\ref{fig: topological partitions}. 
In this specific case, $m^{\ }_{1B} = 2$, with two distinct 
$\Sigma^{\ }_{1\P^{\ }_{1}}$ and $\Sigma^{\ }_{1\P^{\ }_{2}}$ and 
$m^{\ }_{2B} = m^{\ }_{3B} = m^{\ }_{4B} = 1$, with a single 
$\Sigma^{\ }_{2\P}$, $\Sigma^{\ }_{3\P}$, $\Sigma^{\ }_{4\P}$, respectively.

{}From Eq.~(\ref{eq: S^A_VN(T)}), it follows that 
\begin{widetext}
\bea
S^{\ }_{\textrm{topo}}(T) - S^{\textrm{Kitaev}}_{\textrm{topo}} 
&=& 
\sum^{2}_{i=1} 
\left( \tilde{x}^{(1)}_{i} \ln \tilde{x}^{(1)}_{i} \right) 
  \frac{\cosh\left( \frac{K^{\ }_{A}}{2} (N - \Sigma^{\ }_{1\P^{\ }_{i}}) \right)}
       {\cosh\left( \frac{K^{\ }_{A}}{2} N \right)} 
+ 
\sum^{2}_{i=1} 
\left( \tilde{y}^{(1)}_{i} \ln \tilde{y}^{(1)}_{i} \right) 
  \frac{\sinh\left( \frac{K^{\ }_{A}}{2} (N - \Sigma^{\ }_{1\P^{\ }_{i}}) \right)}
       {\cosh\left( \frac{K^{\ }_{A}}{2} N \right)} 
\nonumber \\ 
&& 
- 
\left( \tilde{x}^{(2)}_{\ } \ln \tilde{x}^{(2)}_{\ } \right) 
  \frac{\cosh\left( \frac{K^{\ }_{A}}{2} (N - \Sigma^{\ }_{2\P}) \right)}
       {\cosh\left( \frac{K^{\ }_{A}}{2} N \right)} 
- 
\left( \tilde{y}^{(2)}_{\ } \ln \tilde{y}^{(2)}_{\ } \right) 
  \frac{\sinh\left( \frac{K^{\ }_{A}}{2} (N - \Sigma^{\ }_{2\P}) \right)}
       {\cosh\left( \frac{K^{\ }_{A}}{2} N \right)} 
\nonumber \\ 
&& 
- 
\left( \tilde{x}^{(3)}_{\ } \ln \tilde{x}^{(3)}_{\ } \right) 
  \frac{\cosh\left( \frac{K^{\ }_{A}}{2} (N - \Sigma^{\ }_{3\P}) \right)}
       {\cosh\left( \frac{K^{\ }_{A}}{2} N \right)} 
- 
\left( \tilde{y}^{(3)}_{\ } \ln \tilde{y}^{(3)}_{\ } \right) 
  \frac{\sinh\left( \frac{K^{\ }_{A}}{2} (N - \Sigma^{\ }_{3\P}) \right)}
       {\cosh\left( \frac{K^{\ }_{A}}{2} N \right)}  
\nonumber \\ 
&& 
+
\left( \tilde{x}^{(4)}_{\ } \ln \tilde{x}^{(4)}_{\ } \right) 
  \frac{\cosh\left( \frac{K^{\ }_{A}}{2} (N - \Sigma^{\ }_{4\P}) \right)}
       {\cosh\left( \frac{K^{\ }_{A}}{2} N \right)} 
+ 
\left( \tilde{y}^{(4)}_{\ } \ln \tilde{y}^{(4)}_{\ } \right) 
  \frac{\sinh\left( \frac{K^{\ }_{A}}{2} (N - \Sigma^{\ }_{4\P}) \right)}
       {\cosh\left( \frac{K^{\ }_{A}}{2} N \right)}, 
\label{eq: S_topo(T)}
\eea
\end{widetext} 
where $S^{\textrm{Kitaev}}_{\textrm{topo}}$ is the topological entropy of 
the GS of the original toric code, which obtains from the 
$-\ln (d^{\ }_{\A} d^{\ }_{\B}) / \vert G \vert$ contribution to 
$S^{\A}_{\textrm{VN}}(T)$. All the contributions from the $(x \ln x)$ and 
$(y \ln y)$ terms cancel since 
$
\Sigma^{\ }_{1\A} 
-
\Sigma^{\ }_{2\A} 
-
\Sigma^{\ }_{3\A} 
+ 
\Sigma^{\ }_{4\A} 
= 
0
$ 
by construction. 

For finite systems, the known limiting values are recovered: 
\bea
S^{\ }_{\textrm{topo}}(T \to 0) - S^{\textrm{Kitaev}}_{\textrm{topo}} 
&\rightarrow& 
0 
\label{eq: S_topo(T) T -> infty}
\\ 
S^{\ }_{\textrm{topo}}(T \to \infty) - S^{\textrm{Kitaev}}_{\textrm{topo}} 
&\rightarrow& 
- \ln 2. 
\label{eq: S_topo(T) T -> 0}
\eea

In the thermodynamic limit $L \to \infty$ and with $r,R$ kept constant, 
all $\Sigma^{\ }_{\alpha\P^{\ }_{i}}$ diverge with the exception of 
$\Sigma^{\ }_{1\P^{\ }_{1}}$, which corresponds to the inner square of size 
$(R-2r)^{2}_{\ }$ in Fig.~\ref{fig: topological partitions}. 
Thus, Eq.~(\ref{eq: S_topo(T)}) becomes 
\begin{widetext}
\bea
S^{\ }_{\textrm{topo}}(T) - S^{\textrm{Kitaev}}_{\textrm{topo}} 
&=& 
\sum^{2}_{i=1} 
\left( 
  \tilde{x}^{(1)}_{i} \ln \tilde{x}^{(1)}_{i} 
  + 
  \tilde{y}^{(1)}_{i} \ln \tilde{y}^{(1)}_{i} 
\right) 
  e^{-\frac{K^{\ }_{A}}{2} \Sigma^{\ }_{1\P^{\ }_{i}}}_{\ } 
- 
\left( 
  \tilde{x}^{(2)}_{\ } \ln \tilde{x}^{(2)}_{\ } 
  +
  \tilde{y}^{(2)}_{\ } \ln \tilde{y}^{(2)}_{\ } 
\right) 
  e^{-\frac{K^{\ }_{A}}{2} \Sigma^{\ }_{2\P}}_{\ } 
\nonumber \\ 
&& 
- 
\left( 
  \tilde{x}^{(3)}_{\ } \ln \tilde{x}^{(3)}_{\ } 
  + 
  \tilde{y}^{(3)}_{\ } \ln \tilde{y}^{(3)}_{\ } 
\right) 
  e^{-\frac{K^{\ }_{A}}{2} \Sigma^{\ }_{3\P}}_{\ } 
+
\left( 
  \tilde{x}^{(4)}_{\ } \ln \tilde{x}^{(4)}_{\ } 
  + 
  \tilde{y}^{(4)}_{\ } \ln \tilde{y}^{(4)}_{\ } 
\right) 
  e^{-\frac{K^{\ }_{A}}{2} \Sigma^{\ }_{4\P}}_{\ } 
\nonumber \\ 
&=& 
\left( 
  \tilde{x}^{(1)}_{1} \ln \tilde{x}^{(1)}_{1} 
  + 
  \tilde{y}^{(1)}_{1} \ln \tilde{y}^{(1)}_{1} 
\right) 
  e^{-\frac{K^{\ }_{A}}{2} \Sigma^{\ }_{1\P^{\ }_{1}}}_{\ } 
+ 
\frac{K^{\ }_{A}}{2} \Sigma^{\ }_{1\P^{\ }_{2}} 
- 
\frac{K^{\ }_{A}}{2} \Sigma^{\ }_{2\P}
- 
\frac{K^{\ }_{A}}{2} \Sigma^{\ }_{3\P}
+
\frac{K^{\ }_{A}}{2} \Sigma^{\ }_{4\P} 
\nonumber \\ 
&=& 
\left( 
  \tilde{x}^{(1)}_{1} \ln \tilde{x}^{(1)}_{1} 
  + 
  \tilde{y}^{(1)}_{1} \ln \tilde{y}^{(1)}_{1} 
\right) 
  e^{-\frac{K^{\ }_{A}}{2} \Sigma^{\ }_{1\P^{\ }_{1}}}_{\ } 
- 
\frac{K^{\ }_{A}}{2} \Sigma^{\ }_{1\P^{\ }_{1}}, 
\label{eq: S_topo(T) large L}
\eea
\end{widetext} 
where we used the fact that 
\bea
N 
&=& 
\Sigma^{\ }_{1\A} 
+ 
\Sigma^{\ }_{1\P^{\ }_{1}} 
+ 
\Sigma^{\ }_{1\P^{\ }_{2}}
\\ 
&=& 
\Sigma^{\ }_{2\A} 
+ 
\Sigma^{\ }_{2\P} 
\\
&=& 
\Sigma^{\ }_{3\A} 
+ 
\Sigma^{\ }_{3\P} 
\\
&=& 
\Sigma^{\ }_{4\A^{\ }_{1}} 
+ 
\Sigma^{\ }_{4\A^{\ }_{2}} 
+ 
\Sigma^{\ }_{4\P}, 
\eea 
and that 
\beq
\Sigma^{\ }_{1\A} 
-
\Sigma^{\ }_{2\A} 
-
\Sigma^{\ }_{3\A} 
+ 
\Sigma^{\ }_{4\A} 
= 
0
\eeq
to substitute 
\beq
\Sigma^{\ }_{1\P^{\ }_{2}} 
- 
\Sigma^{\ }_{2\P} 
- 
\Sigma^{\ }_{3\P} 
+ 
\Sigma^{\ }_{4\P} 
= 
- \Sigma^{\ }_{1\P^{\ }_{1}} 
\eeq
into Eq.(\ref{eq: S_topo(T) large L}). 
Considering that we are eventually interested in taking limit 
$r,R \to \infty$ (i.e., $\Sigma^{\ }_{1\P^{\ }_{1}} \gg 1$), 
we obtain the asymptotic value 
\bea
S^{\ }_{\textrm{topo}}(T) - S^{\textrm{Kitaev}}_{\textrm{topo}} 
&\stackrel{L \to \infty}{\longrightarrow}& 
- \ln 2 
\eea
for any non-zero value of $T$, 
that is \emph{thermal equilibrium at any infinitesimal temperature leads to 
a finite loss of topological entropy in the thermodynamic limit}. 
In Sec.~\ref{sec: conclusions} we discuss the implications of this result, 
and in particular, we propose an interpretation that naturally explains 
why the topological entropy reduces to precisely half of its zero-temperature 
value $S^{\textrm{Kitaev}}_{\textrm{topo}} = 2 \ln 2$. 

\begin{figure}[!ht]
\vspace{0.2 cm}
\includegraphics[width=0.98\columnwidth]{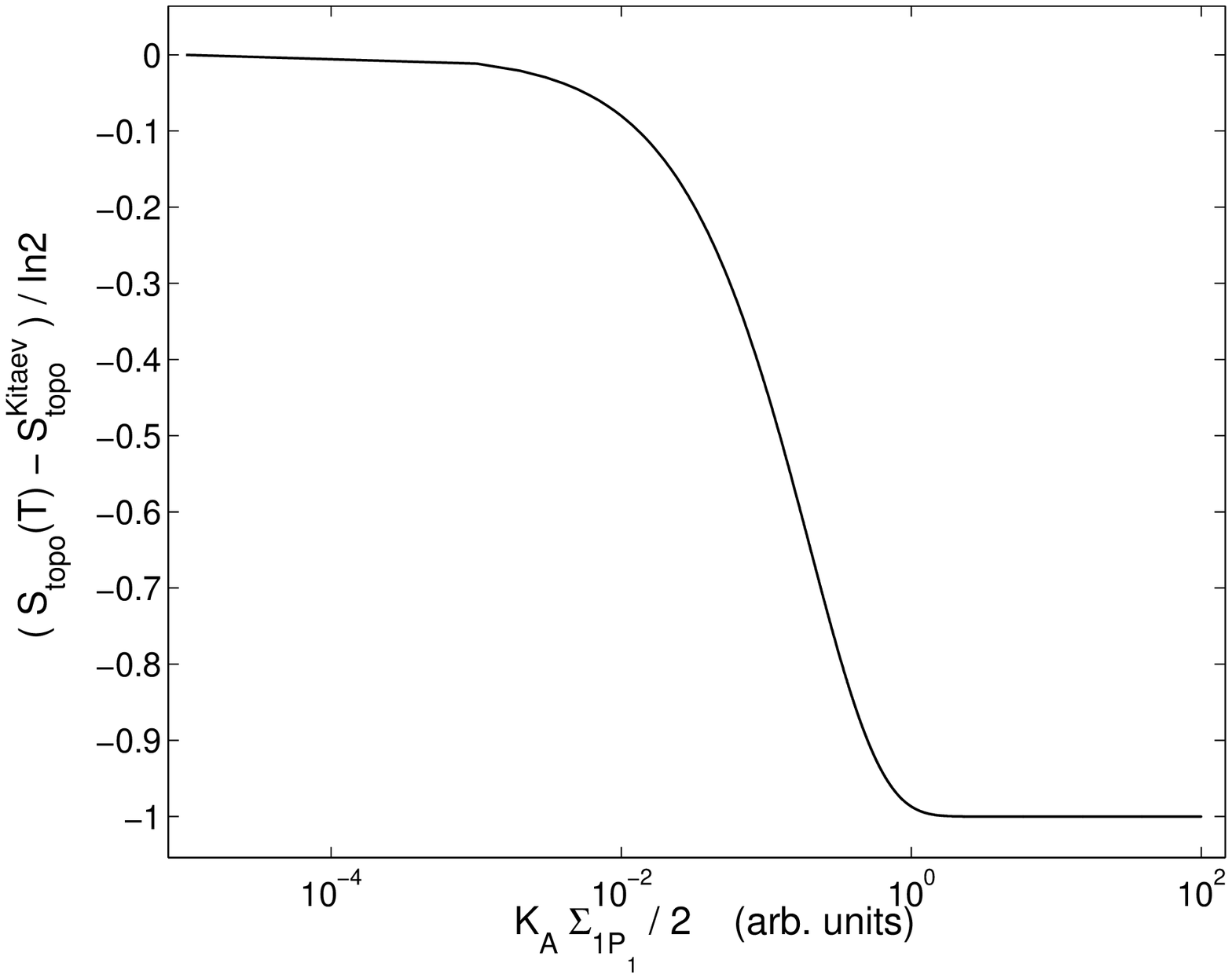}
\\ 
\includegraphics[width=0.98\columnwidth]{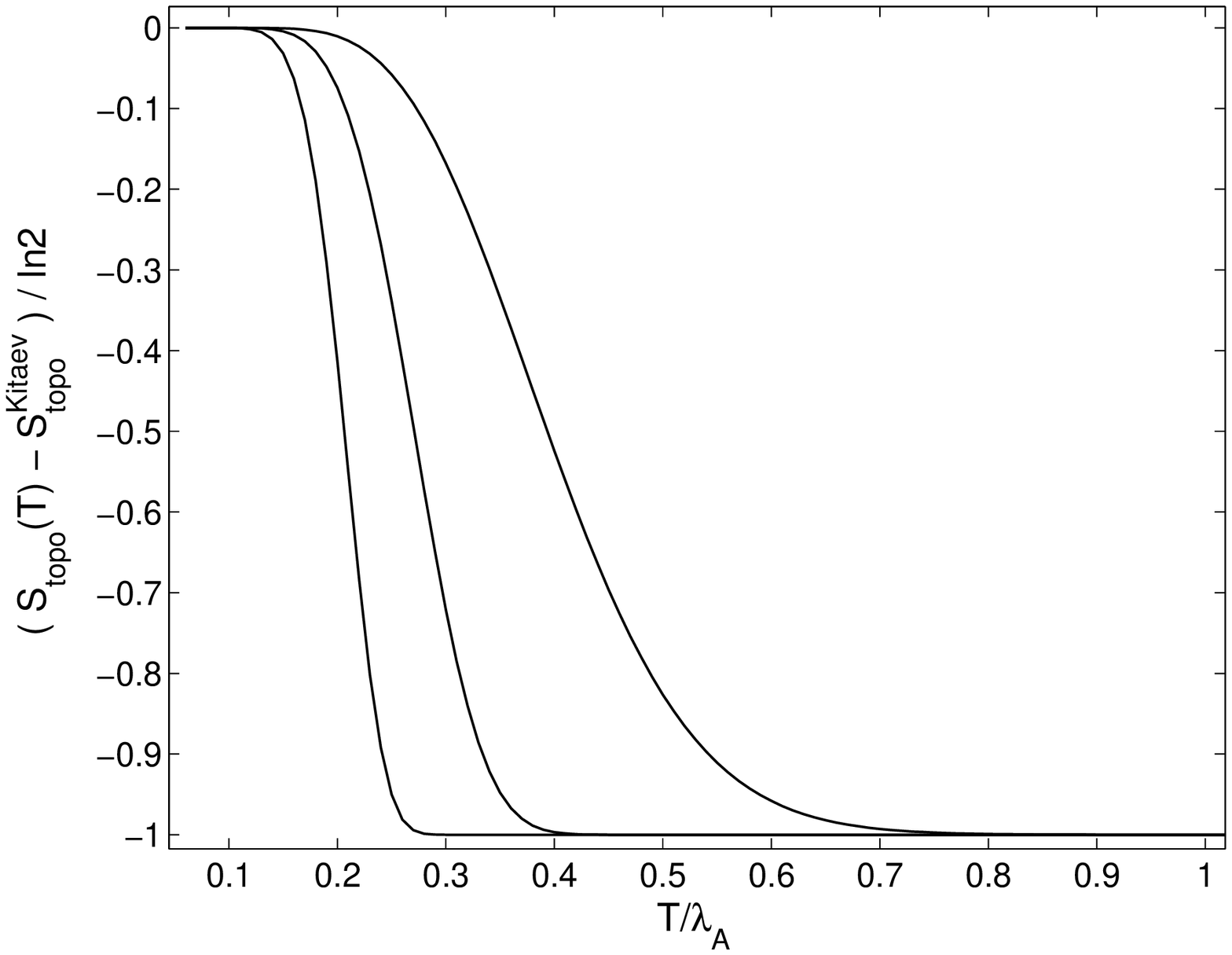}
\caption{
\label{fig: S_topo(chi)}
(Top) 
Limiting behavior of the entropy difference Eq.~(\ref{eq: S_topo(T) large L}) 
in units of $\ln 2$ in the thermodynamic limit, 
as a function of $K^{\ }_{A} \Sigma^{\ }_{1\P^{\ }_{1}} / 2$, 
where $K^{\ }_{A} = -\ln[\tanh(\lambda^{\ }_{A} / T)]$ and 
$\Sigma^{\ }_{1\P^{\ }_{1}} \sim (R-2r)^{2}_{\ }$, the area of the inner 
square in Fig.~\ref{fig: topological partitions}. 
Notice the logarithmic scale on the horizontal axis. 
(Bottom) 
The same curve represented as a function of $T / \lambda^{\ }_{A}$, for 
three different values of $\Sigma^{\ }_{1\P^{\ }_{1}} = 20, 200, 2000$ 
(from right to left). 
}
\end{figure}

Notice that Eq.~(\ref{eq: S_topo(T) large L}) is consistent with both the 
zero-temperature and the infinite-temperature limits, 
Eqs.~(\ref{eq: S_topo(T) T -> infty},\ref{eq: S_topo(T) T -> 0}). 
Notice also that the topological entropy in the $L \to \infty$ limit becomes 
a pure function of $K^{\ }_{A} \Sigma^{\ }_{1\P^{\ }_{1}} / 2$, 
whose shape is illustrated in Fig.~\ref{fig: S_topo(chi)}
The location of the drop, say when 
$
S^{\ }_{\textrm{topo}}(T) 
- 
S^{\textrm{Kitaev}}_{\textrm{topo}} 
= 
- (\ln 2) / 2
$, 
is given by 
\beq
\frac{K^{\ }_{A}}{2} \, \Sigma^{\ }_{1\P^{\ }_{1}} 
\simeq 
\frac{1}{4}. 
\eeq
Even for modest partition sizes with 
$\Sigma^{\ }_{1\P^{\ }_{1}} \gtrsim 100$, the drop occurs at rather small 
temperatures and we can approximate 
\beq
K^{\ }_{A} 
= 
-\ln\left[\tanh\left(\frac{\lambda^{\ }_{A}}{T}\right)\right] 
\simeq 
2 e^{-2\frac{\lambda^{\ }_{A}}{T}}_{\ }. 
\eeq
This in turn gives 
\beq
\Sigma^{\ }_{1\P^{\ }_{1}} 
  e^{-2\frac{\lambda^{\ }_{A}}{T^{\ }_{\textrm{drop}}}}_{\ } 
\simeq 
\frac{1}{4} 
\;\;\;
\Longrightarrow
\;\;\;
T^{\ }_{\textrm{drop}} 
\simeq 
\frac{\lambda^{\ }_{A}}
     {\ln\left( 2 \sqrt{\Sigma^{\ }_{1\P^{\ }_{1}}} \right)}. 
\label{eq: Tdrop Sigma}
\eeq
The l.h.s. of the above equation allows for a straightforward interpretation 
in terms of defects in the underlying electric loop structure. In fact, 
$e^{-\lambda^{\ }_{A} / T}_{\ }$ controls the density of such defects 
in the system, and the equation therefore suggests that the drop in 
topological entropy occurs when the average number of defects inside 
partition $1\B^{\ }_{1}$ becomes of order one. 

In order to understand the behavior of $S^{\ }_{\textrm{topo}}(T)$ at 
finite temperature and finite system size, notice that the temperature 
parameter $K^{\ }_{A} = - \ln [\tanh (\beta\lambda^{\ }_{A})]$ in 
Eq.~(\ref{eq: S_topo(T)}) always appears multiplied by an extensive quantity, 
be it $N$ or one of the $\Sigma$'s. 
It is therefore convenient to make the reasonable assumption 
that the number of star operators in each subsystem 
$\A^{\ }_{1}$, \ldots, $\A^{\ }_{m^{\ }_{\A}}$, 
and  $\B^{\ }_{1}$, \dots, $\B^{\ }_{m^{\ }_{\B}}$ scales linearly with 
the total number of star operators $N$. 
Namely, this amounts to increasing uniformly both $L$ and $r,R$ while keeping 
their ratios fixed, thus simply rescaling the bipartitions in 
Fig.~\ref{fig: topological partitions}. 
We can then introduce the notation 
$\Sigma^{\ }_{\A} = N \gamma^{\ }_{\A}$, 
and 
$\Sigma^{\ }_{\P^{\ }_{i}} = N \gamma^{\ }_{\P^{\ }_{i}}$, 
with $\gamma^{\ }_{\A}, \gamma^{\ }_{\P^{\ }_{i}} \in (0,1)$, 
and $\gamma^{\ }_{\A} + \sum^{\ }_{i} \gamma^{\ }_{\P^{\ }_{i}} = 1$. 
Recalling the definitions of 
$
\tilde{x}^{(\alpha)}_{i} 
= 
\cosh( K^{\ }_{A} \Sigma^{\ }_{\alpha\P^{\ }_{i}} / 2 )
$ 
and 
$
\tilde{y}^{(\alpha)}_{i} 
= 
\sinh( K^{\ }_{A} \Sigma^{\ }_{\alpha\P^{\ }_{i}} / 2 )
$, 
one can replace $K^{\ }_{A}$ by $\mathcal{K}^{\ }_{\A} = K^{\ }_{A} N$ 
and all other parameters in 
Eq.~(\ref{eq: S_topo(T)}) become intensive quantities that do not scale 
with the system size. 
Temperature and system size are strongly 
bound together into a single tunable parameter $\mathcal{K}^{\ }_{\A}$ in 
our system. 
The thermodynamic limit at zero temperature is singular, in that the 
behavior of $\mathcal{K}^{\ }_{\A}$ depends on the order of limits. 
%
%

\subsection{\label{sec: numerics}
Numerical evaluation of $S^{\ }_{\textrm{topo}}(T)$
           }
\begin{figure}[!ht]
\vspace{0.0 cm}
\includegraphics[width=0.98\columnwidth]{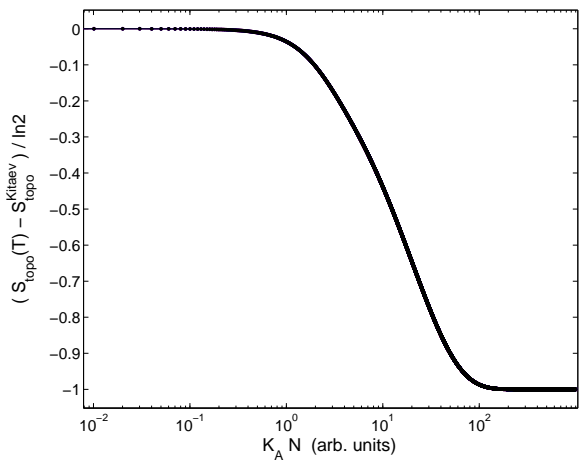}
\caption{
\label{fig: KN collapse}
Topological entropy as a function of $\mathcal{K}^{\ }_{\A}$ for increasing 
system sizes $N = 10^{3}_{\ }, 10^{6}_{\ }, 10^{9}_{\ }$. 
Notice the complete overlap between the different curves, due to the fact 
that the topological entropy Eq.~(\ref{eq: S_topo(T)}) becomes a pure 
function of $\mathcal{K}^{\ }_{\A} = K^{\ }_{A} N$ when all $\Sigma$'s 
scale linearly with $N$. 
Notice the logarithmic scale on the horizontal axis. 
}
\end{figure}

The expression for the topological entropy as a function of 
temperature and system size Eq.~(\ref{eq: S_topo(T)}) is rather lengthy 
and non-transparent. 
In this section we illustrate its behavior graphically, by explicitly 
evaluating $S^{\ }_{\textrm{topo}}(T)$ for small systems. 
In Fig.~\ref{fig: KN collapse} we present the difference 
$(S^{\ }_{\textrm{topo}}(T) - S^{\textrm{Kitaev}}_{\textrm{topo}})/\ln 2$ 
as a function of $\mathcal{K}^{\ }_{\A} = K^{\ }_{A} N$, for various 
system sizes $N = 10^{3}_{\ }, 10^{6}_{\ }, 10^{9}_{\ }$. 
For convenience, we chose the values of $r$ and $R$ proportional to 
$\sqrt{N}$, so that the above assumption on the $\Sigma$'s holds 
true, and $S^{\ }_{\textrm{topo}}(T)$ is a function of 
$\mathcal{K}^{\ }_{\A}$ only. 

In the limit $L \to \infty$, the smooth curves collapse identically onto 
their infinite-temperature value for any non-vanishing temperature, and a 
singularity arises at $T=0$. 

The location of the drop is given by $K^{\ }_{A} N \simeq 10$, from which 
we obtain 
\bea
-\ln\left[\tanh\left(\frac{\lambda^{\ }_{A}}{T^{\ }_{\textrm{drop}}}\right)\right] N 
&\simeq& 
10 
\nonumber \\ 
T^{\ }_{\textrm{drop}} 
&\simeq& 
\frac{\lambda^{\ }_{A}}
     {\tanh^{-1}_{\ }\left(e^{-\frac{10}{N}}_{\ }\right)}. 
\eea
For large enough system sizes, $T^{\ }_{\textrm{drop}}$ is small and the 
above equations reduce to 
\beq
N e^{-2\frac{\lambda^{\ }_{A}}{T^{\ }_{\textrm{drop}}}}_{\ } 
\simeq 
5 
\;\;\; 
\Longrightarrow
\;\;\; 
T^{\ }_{\textrm{drop}} 
\simeq 
\frac{\lambda^{\ }_{A}}
     {\ln\sqrt{N/5}}. 
\label{eq: Tdrop N}
\eeq
Once again, the drop occurs when the average number of defects in the system 
becomes of order one. 
(This is consistent with the previous result in Eq.~(\ref{eq: Tdrop Sigma}) 
since we made here the assumption that all the $\Sigma$'s, and therefore 
$\Sigma^{\ }_{1\P^{\ }_{1}}$ as well, scale linearly with $N$). 
%
%

\section{\label{sec: full temperature range}
The full temperature range
        }
In the regime considered in this paper, finite temperature disrupts
the $\sigma^{\textrm{x}}_{\ }$-loop structure gradually for finite
size systems until it is completely destroyed. This happens while the
$\sigma^{\textrm{z}}_{\ }$-loop structure is fully preserved, and
the topological entropy changes overall from $2 \ln 2$ to 
$\ln 2$ (half of the contribution is lost).

The remaining topological entropy should fade away as temperature is
further increased, and one goes to the regime where defects in the
$\sigma^{\textrm{z}}_{\ }$-loop structure also start to appear, for a
finite energy scale $\lambda^{\ }_{B}$. The temperature scale of the
drop in $S^{\ }_{\textrm{topo}}$ from $\ln 2$ to $0$ corresponds
to when the distance between defects, 
$\xi^{\ }_{B} \sim e^{\lambda^{\ }_{B} /T}$, becomes comparable to the 
system size $L$. 
(Or equivalently, the average number of defects in the system becomes 
roughly of order one -- compare with Eq.~(\ref{eq: Tdrop Sigma}) 
and~(\ref{eq: Tdrop N}).) 

It is not obvious how to obtain the exact expression for this second
step, in contrast with the first step which we calculated exactly in
this paper within the preserved $\sigma^{\textrm{z}}_{\ }$-loop
limit. Nevertheless, we believe that the physical picture is the simple
one (as seen at work in the first drop) that once a handful of defects
appear in that $\sigma^{\textrm{z}}_{\ }$-loop structure, the
topological entropy will plunge much like in the first drop. Pasting the
two pictures together, we have the two-stage drop of the topological
entropy sketched in Fig.~\ref{fig: S_topo vs T full}. 
Clearly, in the limit 
$\vert \lambda^{\ }_{A} - \lambda^{\ }_{B} \vert \to 0$ 
the two drops are expected to merge together, and in particular in the 
thermodynamic limit the topological entropy entirely vanishes for any 
infinitesimal temperature. 

We would like to point out that a notion of fragility in the Kitaev
model at finite temperature, in terms of expectation values of toric
operators, has been discussed by Nussinov and
Ortiz~\cite{Nussinov2006} within their definition of topological
quantum order (based on gauge-like symmetries).

%
%

\section{\label{sec: conclusions}
Conclusions
        }
We calculated the entanglement entropy exactly for the toric code at
finite temperatures, in a regime where there is a broad separation of
energy scales between the two couplings in the problem, $\lambda^{\
}_{A}\ll \lambda^{\ }_{B}$. These couplings, from a $\mathbb Z_2$
gauge theory perspective, correspond to the chemical potentials of electric 
charges and magnetic monopoles. One can define length scales associated with
the separation between these types of defects, $\xi^{\ }_{A,B}\sim
e^{\lambda^{\ }_{A,B} /T}$, and for system sizes much smaller than the 
largest of these two length scales, i.e., $L\ll \xi^{\ }_{B}$, 
one of the two loop structures in
the system, associated with the $\sigma^{\textrm{z}}_{\ }$-basis, is
preserved. This is the regime where magnetic monopoles are not present
in the finite size system. In the limit $\lambda^{\ }_{B}\to\infty$, this 
holds true for any system size. It is in this limit that we obtain the
exact result for the entanglement entropy as a function of
$T/\lambda^{\ }_{A}$.

Within this hard constrained regime, we find that the entanglement
entropy is a singular function of temperature and system size, and
that the limit of zero temperature and the limit of infinite system
size do not commute. The two limits differ by a term that does not
depend on the size of the boundary between the partitions of the
system into two entangled parts, but instead depends on the topology
of the bipartition. We also calculate the mutual information, obtained
from the von Neumann entropy by a symmetrization procedure to filter
bulk terms at non-zero temperatures and to leave only boundary and
topological contributions. Similarly, the difference between the two
orders of limits is an ${\cal O}(1)$ term that is purely topological,
depending on the number of disconnected pieces of partitions $\A$ and
$\B$.

We find that one half of the topological entropy is shaved off
from its $T=0$ value as the temperature increases above $T^{(A)}_{\rm
cross} \sim \lambda^{\ }_{A} / \ln\sqrt{N}$. Above this scale, the
loop structure associated with the $\sigma^{\textrm{x}}_{\ }$-basis is
destroyed, while the one associated with the $\sigma^{\textrm{z}}_{\
}$-basis survives (recall the $\lambda^{\ }_{B}\to\infty$). 
We argue that a large but finite value of $\lambda^{\ }_{B}$ would 
introduce another scale 
$T^{(B)}_{\rm cross} \sim \lambda^{\ }_{B} / \ln\sqrt{N}$, above 
which the rest of the topological entropy should also vanish.

As these results show, the topological contributions to the von Neumann
entropy or equivalently to the topological entropy, are rather fragile 
for non-zero temperatures. If the thermodynamic limit is taken first,
these quantities subside immediately. However, in practice one should
focus on physical regimes and not mathematical limits. The reason why
these quantities are so fragile is that ${\cal O}(1)$ defects can
destroy them. However, one must realize that the length scale
associated to the defect separation grows exponentially as temperature
is decreased, and becomes astronomical for temperatures a 
few hundred times smaller than the energy scales $\lambda^{\
}_{A,B}$. Hence, even if these topological contributions to the
entanglement entropy technically vanish, they are statistically present 
in large but laboratory size physical systems. 

If one is interested in understanding how robust is the topological 
order (information) stored in a single finite system, the notion of a 
statistically non-vanishing topological entropy naturally translates 
into the presence of a characteristic time scale over which 
topoogical order is preserved. 
Such time scale is associated with the Boltzman probability for the 
appearance of a defect, 
namely $\mathcal{N}\,e^{-\lambda^{\ }_{A} / T}_{\ }$, 
where $\mathcal{N}$ is the total number of degrees of freedom in the 
system. 
In sight of a possible practical use of such topological quantum 
information, it would therefore be of great importance to compare 
this persistence time scale with the one associated to the preparation 
of the system into a topologically ordered state. 
Preliminary research in that direction can be found in 
Ref.~\onlinecite{Hamma2006} and in Ref.~\onlinecite{Alicki2007}. 

At a more fundamental level, our results suggest a simple pictorial
interpretation of quantum topological order, at least for systems
where there is an easy identification of loop structures as in the
case here studied. Recall that we start from a zero-temperature system
exhibiting quantum topological order associated with the presence of
two identical underlying closed-loop structures. In particular, the
corresponding topological entropy equals $\ln D^{2}_{\ }$, where $D=2$
is the so-called quantum dimension of the system. By allowing one of
the two loop structures to be thermally disrupted, and by raising $T
\to \infty$ while the other loop structure is fully preserved, we arrive 
at a classical system with a single (therefore classical) underlying loop
structure, and exhibiting precisely half of the original topological
entropy ($\ln D$). This is strongly suggestive that
(i) the two loops structures contribute equally and independently to the 
topological order at zero temperature; 
(ii) each loop structure \emph{per se} is a classical (non-local) object 
carrying a contribution of $\ln D$ to the topological entropy; 
and (iii) the quantum nature of the zero-temperature system resides in
the fact that two independent loop structures are allowed to be
superimposed and thus coexist in the system.
In this sense, our results lead to an interpretation of quantum
topological order, at least for systems with simple loop or membrane 
structures, as the quantum mechanical version of a classical topological 
order (given by each individual loop structure). 

Finally, we would like to comment on the fact that the same ${\cal
O}(1)$ defects that deteriorate the topological entropy of the system
should also deteriorate its usefulness for topological quantum
computing. A handful of stray unaccounted defects winding and braiding
around others that are accounted for in the computational scheme will
lead to errors. These defects can be thermally suppressed, if the
temperature is small enough and the system not too large, 
so that unwanted defects have a small probability of appearing
in the sample. Thus, quantifying topological entropy at finite
temperature and finite system size is meaningful in quantifying, in a
statistical sense, the degree with which a physical (finite) system 
retains topological order. 

Although the results presented here were derived in the case of one of the 
coupling constants being infinite, we have recently been able to extend the 
calculations to the case where both coupling constants are finite 
[C. Castelnovo and C. Chamon, in preparation]. 
The two contributions to the topological entropy due to the underlying gauge 
structures are shown to behave additively, and indeed the behavior 
conjectured in Fig.~\ref{fig: S_topo vs T full} is confirmed. 
%
%

\section*{
Acknowledgments
         }
We are grateful to Xiao-Gang Wen for his insightful comments on the 
loop structure underlying our model, and to Eduardo Fradkin for enlightening 
discussions. 
This work is supported in part by the NSF Grant DMR-0305482 (C.~Chamon), 
and by EPSRC Grant No. GR/R83712/01 (C.~Castelnovo). 
C.~Castelnovo would like to acknowledge the I2CAM NSF Grant DMR No. 0645461 
for travel support, during which part of this work was carried out. 
%
%


\begin{thebibliography}{99}
\bibitem{Wen-comment} 
We are indebted to Xiao-Gang Wen for pointing out to us the possibility of 
a much richer behavior in higher dimensions, due to the different nature of 
the loop and membrane structures underlying topological order.
\bibitem{footnote: large partitions}
Notice that, while the constraint 
$\Sigma^{\ }_{\A} + \sum^{\ }_{i} \Sigma^{\ }_{\P^{\ }_{i}} = N$ 
requires at least one of the $\Sigma$'s to diverge for $L \to \infty$, 
this does not need to be the case for all of them, in general.
\end{thebibliography}
\end{document}